\documentclass[10pt,twocolumn]{article}

\usepackage{times}
\usepackage{fullpage}
\usepackage{url}
\usepackage{graphicx}
\usepackage{epstopdf}
\usepackage{subfigure}
\usepackage{multirow}
\usepackage{algorithm}
\usepackage[noend]{algpseudocode}
\usepackage{amsmath}
\usepackage{amsfonts}
\usepackage[colorlinks,
            linkcolor=red,
            anchorcolor=blue,
            citecolor=green
            ]{hyperref}

\newtheorem{thm}{Theorem}

\newtheorem{defn}{Definition}

\begin{document}

\title{\bf Towards \textit{Reliable} (and \textit{Efficient}) Job Executions in a \textit{Practical} Geo-distributed Data Analytics System}
\author{Xiaoda Zhang, Zhuzhong Qian, Sheng Zhang, Yize Li, Xiangbo Li, Xiaoliang Wang, Sanglu Lu \\
\textit{State Key Laboratory for Novel Software Technology, Nanjing University}}
\date{}
\maketitle
\thispagestyle{empty}

\begin{abstract}
Geo-distributed data analytics are increasingly common to derive useful information in large organisations. Naive extension of existing cluster-scale data analytics systems
to the scale of geo-distributed data centers faces unique challenges including WAN bandwidth limits, regulatory constraints, changeable/unreliable runtime environment, and high monetary costs.
Our goal in this work is to develop a practical geo-distributed data analytics system that (1) employs an intelligent mechanism for jobs to efficiently utilize (adjust to) the resources (changeable environment) across data centers; (2)  guarantees the reliability of jobs due to the possible failures; and (3) is generic and flexible enough to run a wide range of data analytics jobs without requiring any changes.

To this end, we present a new, general geo-distributed data analytics system, \textsc{Houtu}, that is composed of multiple autonomous systems, each operating in a sovereign data center. \textsc{Houtu} maintains a job manager (JM) for a geo-distributed job in each data center, so that these replicated JMs could \textit{individually} and \textit{cooperatively} manage resources and assign tasks. Our experiments on the prototype of \textsc{Houtu} running across four Alibaba Cloud regions show that \textsc{Houtu} provides efficient job performance as in the existing centralized architecture, and guarantees reliable job executions when facing failures.
\end{abstract}

\section{Introduction}

Nowadays, organizations are deploying their applications in multiple data centers around the world to meet the latency-sensitive requirements~\cite{AWS, GoogleDC, AzureDC, Aliyun}. As a result, the raw data -- including user interaction logging, compute infrastructure monitoring, and job traces -- is generated at geographically distributed data centers. Analytics jobs on these geo-distributed data are emerging as a daily requirement \cite{Mesa, JetStream, Regulatory, SPANStore, Hung, Pu, CLARINET, Lube, Gaia, baochun}.

Because these analytics jobs usually support the real-time decisions and online predictions, minimizing response time and maximizing throughput are important.  However, these face the unique challenges of wide area network (WAN) bandwidth limits, legislative and regulatory constraints, unreliable runtime environment, and even monetary costs.

\begin{figure}
  \centering
  \includegraphics[width=2.9in]{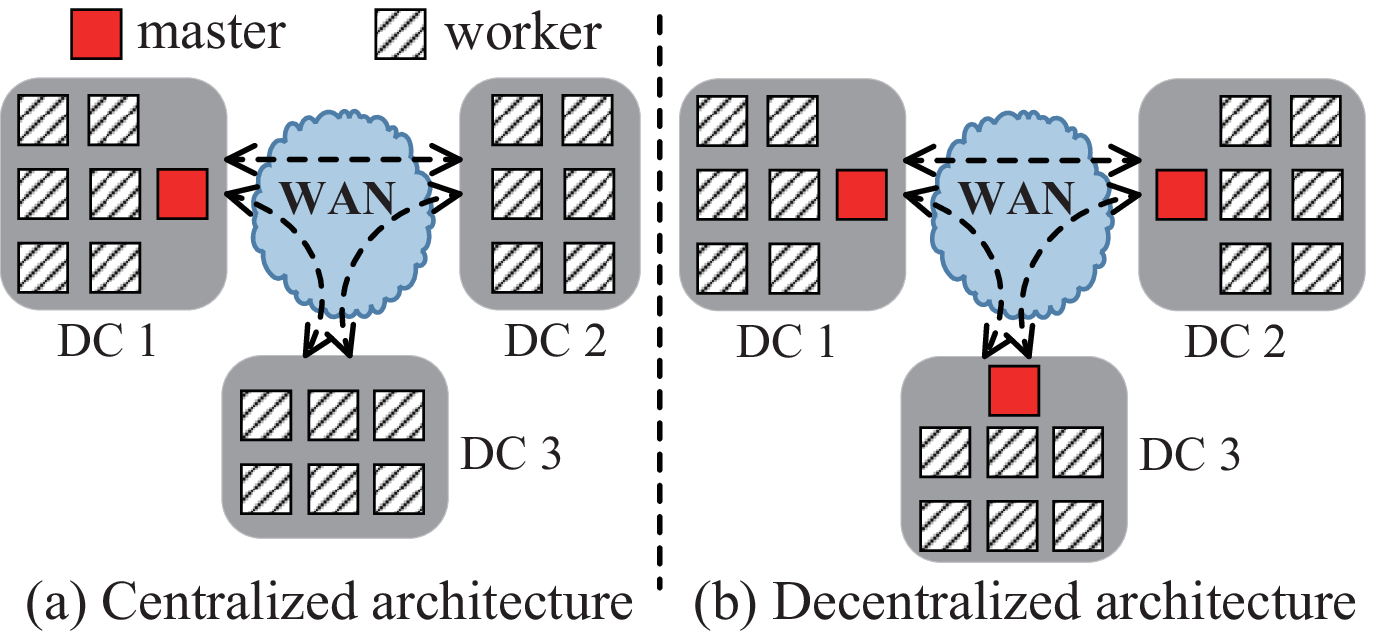}\\
  \caption{Centralized vs. decentralized data analytics.}\label{figure1}
\end{figure}

Existing approaches optimize tasks and/or data placement across data centers so as to improve data locality \cite{PIXIDA, Regulatory, Hung, Pu, CLARINET, Lube}. However, all previous works employ a centralized architecture where a monolithic master controls the resources of the worker machines from  all data centers, as shown in Fig.~\ref{figure1}(a). We argue that regulatory constraints prevent us to do so. More and more regions are establishing laws to restrict the data movement \cite{EU, Dud, HK} and to restrict IT resources from being controlled by other untrusted parties in the shared environment \cite{buyya} (\S\ref{regulatory}). An alternative way is to deploy an \textit{autonomous} data analytics system per data center (Fig.~\ref{figure1}(b)), and extend the original system functionalities to coordinate for geo-distributed job executions. We explore this decentralized architecture and its potentialities, making it possible for a job to acquire resources from remote data centers which respects to the regulatory constraints.

In addition, most existing works assume that the WAN bandwidth is stable. This may not accurately conform to the reality \cite{Hong, Jain}, and our experiments verify that data transmission rate across data centers \textit{varies} even in a short period (\S\ref{changeable}). Hence, this restriction does not allow us to explicitly formulate WAN bandwidth as a constant.

On the other hand, for most organizations who have the geo-distributed data analytics requirement, the most convenient way is to purchase public cloud instances. Decisions must be made between choosing reliable (Reserved and On-demand) instances and unreliable (Spot) instances, due to the different monetary costs and job reliability demands. Spot market  prices are often significantly lower -- by up to an order of magnitude -- than fixed prices for the same instances with a reliability Service Level Agreement (SLA) (\S\ref{spot}). However, is it possible for cloud users to obtain reliability from unreliable instances with a reduced cost? There are positive answers by designing user bidding mechanisms \cite{Wolski, Zheng}, while we answer this question in a systematic way, by providing job-level fault tolerance.

Our goal in this new decentralized and changeable/unreliable environment is to design new resource management, task scheduling and fault tolerance strategies to achieve reliable and efficient job executions.

To achieve this goal, such a system needs to address three key challenges. First, we need to find an efficient scheduling strategy that can dynamically adapt scheduling decisions to the changeable environment. This is difficult because we do not assume job characteristics as a priori knowledge \cite{corral}, or use offline analysis \cite{Ernest} for its significant overhead. Second, we need to implement fault tolerance mechanism for jobs running atop unreliable Spot instances. Though existing frameworks  \cite{mapreduce, Dryad, Spark} tolerate task-level failures, the job-level fault tolerance is absent. While in the unreliable setting, the two types of failures have the same chance to occur. Third, we need to design a \textit{general} system that efficiently handles geo-distributed job executions without requiring any job description changes. This is challenging because data can disperse among sovereign domains (data centers) with regulatory constraints.

In this work, we present \textsc{Houtu}\footnote{\textsc{Houtu} is the deity of deep earth in ancient Chinese mythology who controls lands from all regions.}, a new general geo-distributed data analytics system that is designed to efficiently operate over a collection of data centers.  The key idea of \textsc{Houtu} is to maintain a job manager (JM) for the geo-distributed job in each data center, and each JM can \textit{individually} assign tasks within its own data center, and also \textit{cooperatively} assign tasks between data centers. This differentiation allows \textsc{Houtu} to run conventional task assignment algorithms within a data center \cite{DelaySched, corral, xiaoda}. At the same time, across different data centers, \textsc{Houtu} employs a new work stealing method, converting the task steals to node update events which respects to the data locality constraints.

For resource management, we classify three cases where each job manager \textit{independently} either requests more resources, or maintains current resources, or proactively releases some resources. The key insight here is using nearly past resource utilization as feedback, irrespectively of the prediction of future job characteristics. Even without the future job characteristics, when cooperating with our new task assignment method, we theoretically prove (under some conditions) the efficiency of job executions by extending the very recent result \cite{xiaoda} (\S\ref{analysis}).

Each replicated JM keeps track of the current process of the job execution. We carefully design what need to be included in the intermediate information, which can be used to successfully recover the failure, of even the primary JM.

We build \textsc{Houtu} in Spark \cite{Spark} on YARN \cite{YARN} system, and leverage Zookeeper \cite{zk} to guarantee the intermediate information consistent among job managers in different data centers.
We deploy \textsc{Houtu} across four regions on Alibaba Cloud (AliCloud). Our evaluation with typical workloads including TPC-H and machine learning algorithms shows that, \textsc{Houtu}: (1) achieves efficient job performance as in the centralized architecture; (2) guarantees reliable job executions when facing job failures; and (3) is very effective in reducing monetary costs.

We make three major contributions:
\begin{itemize}
  \item We present a general decentralized data analytics system to respect the possible regulatory constraints and changeable/unrealible runtime environment. The key idea is to provide a job manager for a geo-distributed job in each data center. The system is general and flexible enough to deploy a wide range of data analytics jobs while requiring no change to the jobs themselves (\S\ref{architecture}).

  \item We propose resource management strategy Af for each JM which  exploits resource utilization as feedback. We design task assignment method Parades which combines the assignment within and between data centers. We prove Af + Parades guarantees efficiency for geo-distributed jobs with respect to makespan (\S\ref{design}). We carefully design the mechanism of coordinating JMs, and the intermediate information to recover a failure (\S\ref{ajob}).

  \item We build a prototype of our proposed system using Spark, YARN and Zookeeper as blocks, and demonstrate its efficiencies over four geo-distributed regions with typical diverse workloads (\S\ref{implementation} and \S\ref{evaluation}). We show that \textsc{Houtu} provides efficient and reliable job executions, and significantly reduces the costs for running these jobs.
\end{itemize}
\section{Background and Motivation}\label{background}
This section motivates and provides background for \textsc{Houtu}. \S\ref{regulatory} describes the existing and upcoming regulatory constraints which prevent us from employing a centralized architecture. We measure the scarce and changeable WAN bandwidth between AliCloud regions in \S\ref{changeable}. We investigate a way to reduce monetary cost using Spot instances in \S\ref{spot}, which introduce the unreliability.

\subsection{Regulatory constraints}\label{regulatory}
Though it is efficient to employ data analytics systems in clouds, many organisations still decline to widely adopt cloud services due to severe confidentiality and privacy concerns \cite{FMSE}, and explicit regulations in certain sectors (healthcare
and finance) \cite{hipaa}. Local governments start to impose constraints on raw data storage and movement \cite{EU, HK, Dud}. These constraints exclude the solutions that move arbitrary raw data between data centers \cite{Pu, CLARINET}.

Public clouds allow users to instantiate virtual machines (instances) on demand. In turn, the use of virtualization allows third-party cloud providers to maximize the utilization of their sunk capital costs by multiplexing many customer VMs across a shared physical infrastructure. However, this approach introduces new \textit{vulnerabilities}. It is possible to map the internal cloud infrastructure, identify
where a particular target VM is likely to reside, and then
instantiate new VMs until one is placed co-resident with the
target, which can then be used to mount cross-VM side-channel attacks to extract information from a target VM \cite{CCS, buyya}. The attack amplifier turns this initial compromise of a host into a platform for launching a broad, cloud-wide attack \cite{CapNet}.

Hence, cloud providers and exiting works are proposing solutions in which a group of instances have their \textit{external} connectivity restricted according to a declared policy as a defense against information leakage \cite{Zhai, iam, vpc}. As a result, these upcoming regulatory constraints lead to deploying an autonomous system in each data center, which contains a \textit{complete} stack of data analytics software.

By following exactly this guideline, we propose a decentralized architecture (Fig.~\ref{figure1}(b)) and design how resource management and task scheduling should be performed to support geo-distributed job executions. We speculate that \textit{derived} information, such
as aggregates and reports (which are critical for business intelligence but have less dramatic privacy
implications) may still be allowed to cross geographical boundaries.

\subsection{Changeable environment}\label{changeable}
\begin{figure}
  \centering
  \includegraphics[width=2.3in]{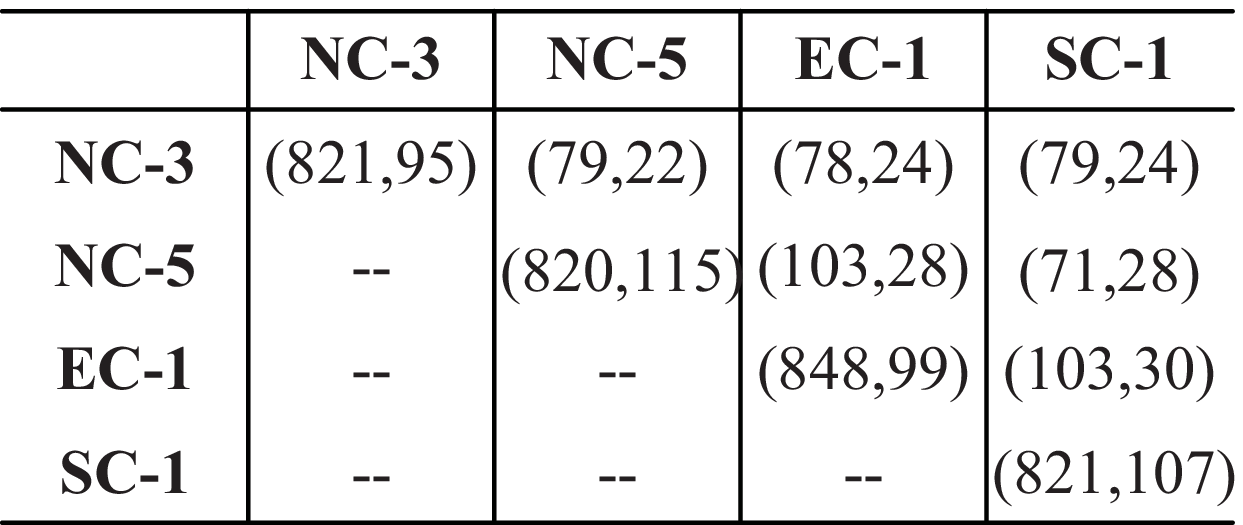}\\
  \caption{Measured network bandwidth between four different regions in AliCloud. The entry is of form (Average, Standard deviation) Mbps.}\label{bandwidth}
\end{figure}

It is well known that WAN bandwidth is a very scarce resource relative to LAN bandwidth.
To quantify WAN bandwidth between data centers, we measure the network bandwidth between all pairs of AliCloud in four regions including NorthChina-3 (NC-3), NorthChina-5 (NC-5), EastChina (EC-1), and SouthChina-1 (SC-1). We measure the network bandwidth of each pair of different regions for three rounds, each for $5$ minutes. As shown in Fig.~\ref{bandwidth}, the bandwidth within a data center is around $820$ Mbps, while around $100$ Mbps between data centers.

What we emphasize is that the WAN bandwidth \textit{varies} between different regions even in a small period. The standard deviation can be as much as 30\% of the available WAN bandwidth itself. The fluctuated bandwidth leads to data transmission time unpredictable \cite{Hong, Jain}.

Furthermore, it may not always be the same resource -- WAN bandwidth -- that causes runtime performance bottlenecks in wide-area data analytics queries. It is confirmed that memory may also becomes the bottleneck at runtime \cite{Lube}, thus these uncertainties do not allow us to assume the capacities of resources (\textit{e.g.} network, compute) as constant in mathematical programming \cite{Pu, CLARINET, Regulatory}. We design intelligent mechanisms that can make online scheduling decisions to the changeable environment.

\subsection{Spot instance: towards reducing cost}\label{spot}
\begin{figure}
  \centering
  \includegraphics[width=2.8in]{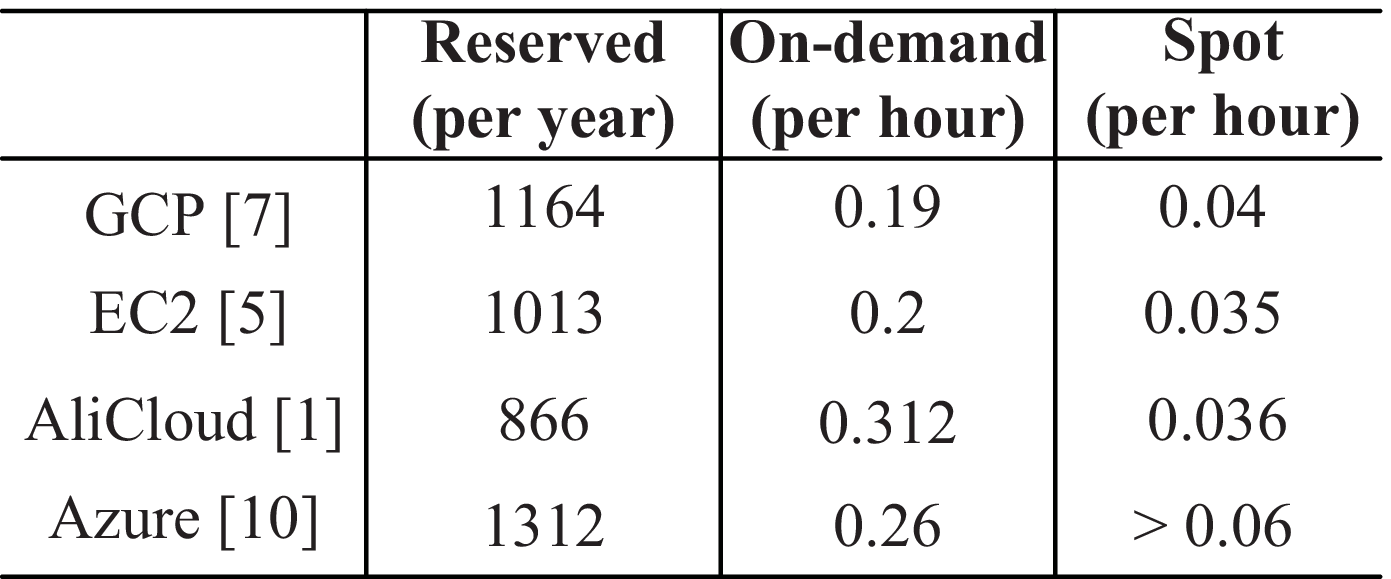}\\
  \caption{Three pricing ways to pay for an instance with $\langle$4 vCPU, 16 GB memory$\rangle$ in GCP, EC2, AliCloud and Azure (in USD).}\label{insprices}
\end{figure}

Cloud computing providers may offer different SLAs at different
prices so that users can control the value transaction at a
fine level of granularity. Besides offering reliable (Reserved and On-demand) instances, cloud providers such as Google Cloud Platform (GCP) \cite{gcp}, Amazon EC2 \cite{awsprice}, Microsoft Azure \cite{azure} and Alibaba Cloud \cite{aliyunprice} also offer ``Spot instances''\footnote{We use this term from EC2, while it is called ``preemptible VM'' in GCP, ``bidding instance'' in AliCloud, and ``low-priority VM'' in Azure.} where resources (at a cheaper price) without a reliability SLA. When a user makes a request for a Spot instance having a specific set of characteristics (\textit{e.g.} $\langle$4 vCPU, 16 GB memory$\rangle$), he/she includes a maximum bid price indicating the maximum that the user is willing to be charged for the instance. Cloud providers create a market for each instance type and satisfy the request of the highest bidders. Periodically, cloud providers recalculate the market price and \textit{terminate} those instances whose maximum bid is below the new market price. Because
the Spot instance market mechanism does not provide a way to guarantee how long an instance will run before it is terminated as part of a SLA, Spot market prices are often significantly lower than fixed prices for the same instances with a reliability SLA (by up to 10x lower than On-demand price, and 3x lower than Reserved price, as shown in Fig.~\ref{insprices}).

Is it possible to deploy a data analytics system using Spot instances and guarantee reliable job executions with reduced cost? To answer this question, it requires to tolerate job-level and task-level failures due to the terminations of unreliable instances, where the former one relates to the failure of job managers. Because both job managers and tasks run in \textit{unified containers}, the two types of failures have the same opportunity to occur. Unfortunately, while the task-level fault-tolerance is implemented in current systems \cite{mapreduce, Spark, Dryad}, these systems do not tolerate job manager failures except for restarting them.

We propose \textsc{Houtu}, which extends the current system functionalties to implement the job-level fault-tolerance, and applies our dynamic scheduling schemes in resource management and task assignment in the decentralized architecture. We experimentally verify the effectiveness and efficiency of \textsc{Houtu}.

\section{System Overview}\label{systemoverview}
We first provide an overview of the \textsc{Houtu} architecture and a job's lifecycle in \textsc{Houtu}. Next we elaborate how a job acts in normal operation and failure recovery.
\begin{figure*}
\centering
\subfigure[\textsc{Houtu}'s architecture and a job's lifecycle.]{
\begin{minipage}[b]{0.44\textwidth}
\includegraphics[width=1\textwidth]{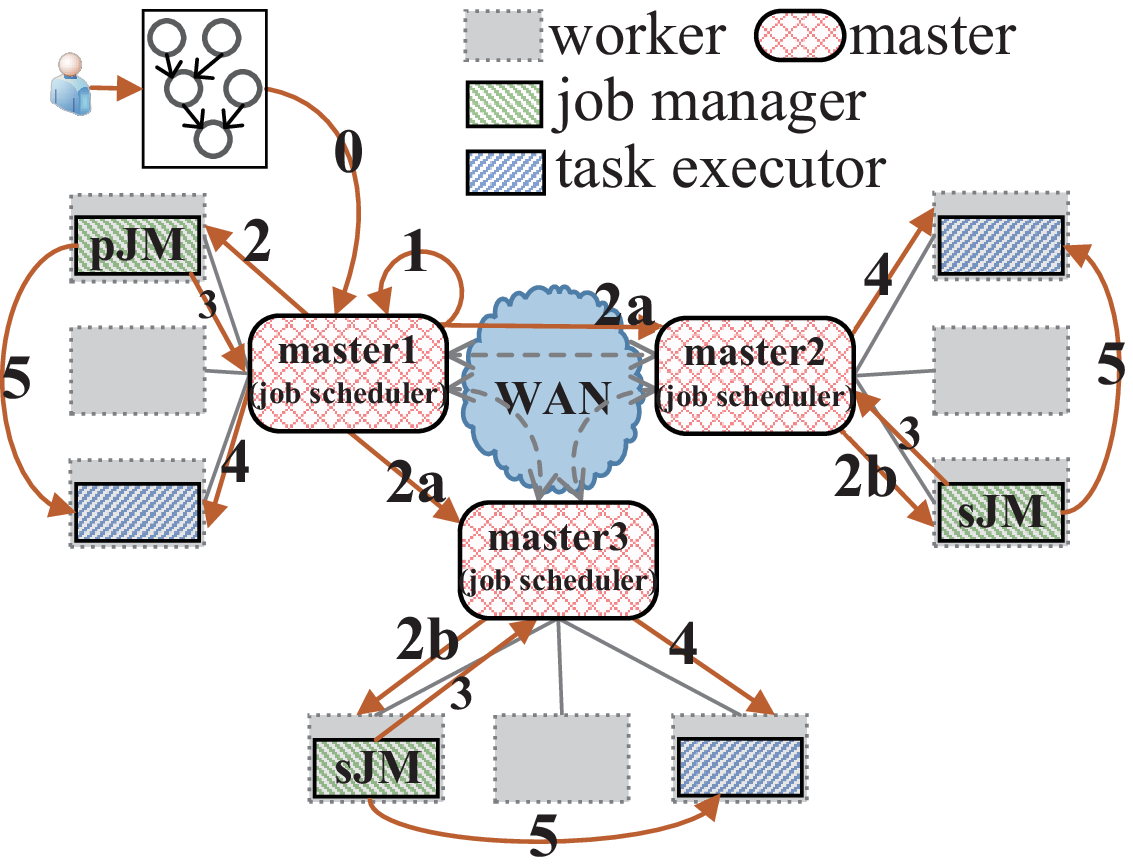}
\end{minipage}
}
\subfigure[A job's \textit{logical} topology in \textsc{Houtu}.]{
\begin{minipage}[b]{0.44\textwidth}
\includegraphics[width=1\textwidth]{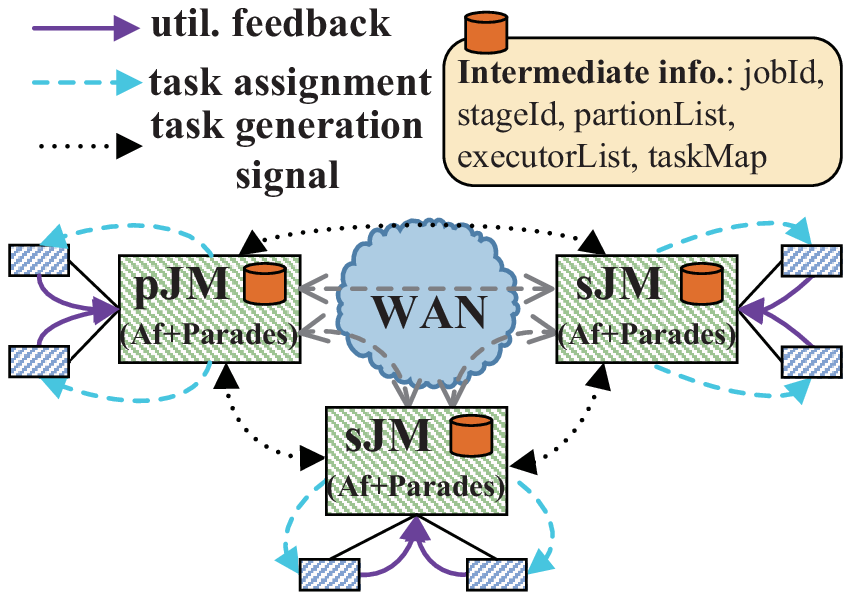}
\end{minipage}
}
\caption{\textsc{Houtu}'s architecture, a job's lifecycle and a job's logical topology in it.}\label{arch}
\end{figure*}

\subsection{\textsc{Houtu} architecture}\label{architecture}
As shown in Fig.~\ref{arch}(a), \textsc{Houtu} is of the decentralized architecture, which is composed with several autonomous systems, deployed in geographically distributed data centers. Each system has the ability to run conventional single-cluster jobs, and also to cooperate with each other to support geo-distributed job executions, while we focus the latter in this work.

\begin{figure}
  \centering
  \includegraphics[width=3in]{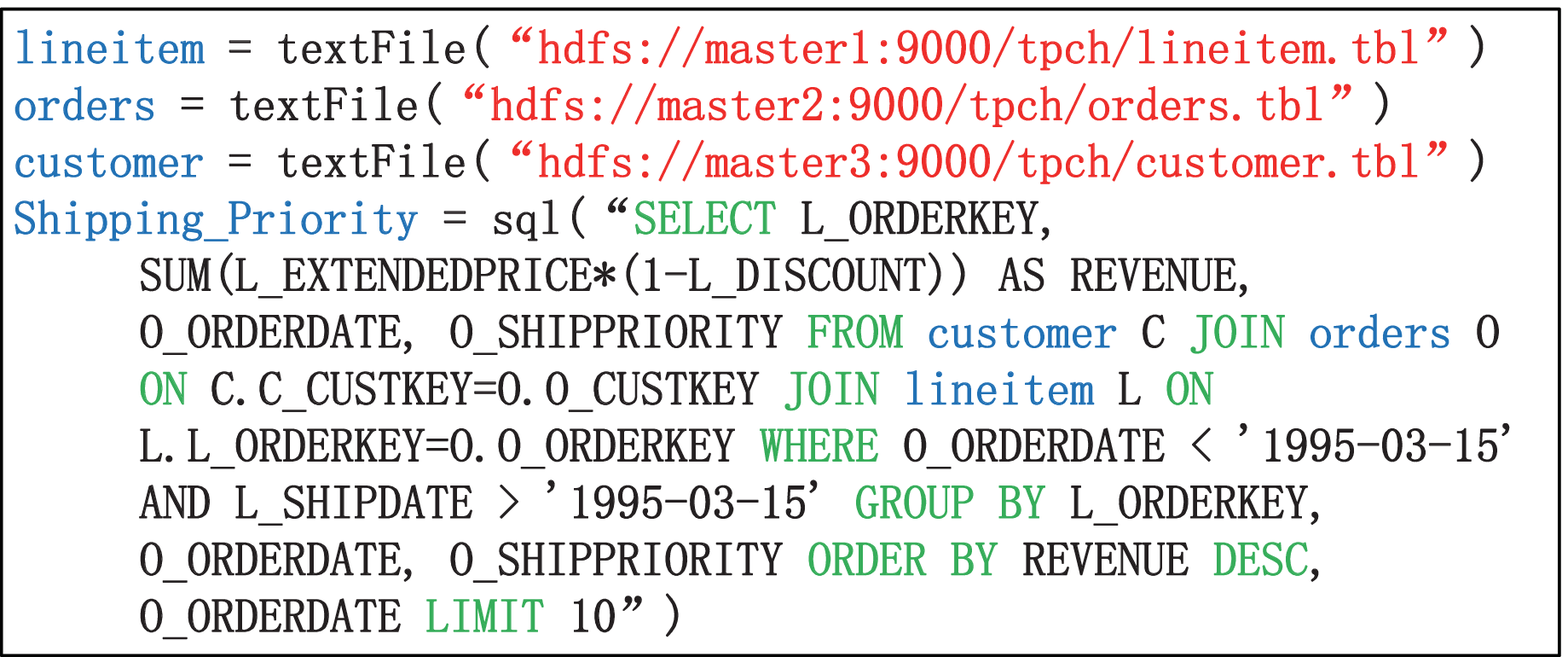}\\
  \caption{Pseudo-code of a job's description in \textsc{Houtu}.}\label{job-descri}
\end{figure}

As stated in \S 1, \textsc{Houtu} is a general system that
efficiently handles geo-distributed job executions without
requiring any job description changes. We speculate that users have the knowledge of how data is distributed across several data centers. The users specify the data locations ``as if'' in a centralized architecture, except with different ``masters''. In the SQL example of Fig.~\ref{job-descri}, three tables are in different data centers, and the job derives statical information from all these tables. \textsc{Houtu} will automatically support the execution of a job described in this way.

Next, we present a job's lifecycle, following the steps of Fig.~\ref{arch}(a).

\textbf{The job submission and job manager generations:} Suppose a user submits the DAG job to a chosen master (step 0). We use
DAG to refer to a directed acyclic graph, where each vertex represents a task and edges encode input-output dependencies. The master would resolve
the job description and generate corresponding job managers for it (step 1). It directly generates a \textit{primary} job manager (pJM) within its own cluster (step 2). For
the remote resources, the master forwards the job description to the remote masters (step 2a) and tells them to generate \textit{semi-active} job manager (sJM) for it (step 2b) (\S\ref{ajob}).

\textbf{Resource request and task executions:} Further, to obtain compute resources (task executors\footnote{Unless otherwise specified, we use the term ``container'' and ``executor'' interchangeably.}), the job managers independently send the requests to their local masters (step 3). The masters (job schedulers) schedule resources to the JMs according to their scheduling invariants, and signal this by returning to the JMs containers that grant access to such resources (step 4). After that, the JMs send the tasks to run in the containers (step 5). As the DAG job is dynamically unfolded and resource requests of the job usually are  not satisfied in a single wave, JMs often repeat steps 3 -- 5 for multiple times.

We leave the design of how job managers request resources without further
characteristics of the unfolding DAG, and how to schedule tasks within a data center and between data centers in \S\ref{design}.
\subsection{A job in \textsc{Houtu}}\label{ajob}
We show how the primary job manager (pJM) and semi-active job managers (sJMs) coordinate to execute a job.
\subsubsection{Normal operation}\label{normaloperation}
In normal operation, there is exactly one pJM and all of the other JMs are sJMs in a job. When the master (to which a user submit the job) forwards the request (step 2a in Fig.~\ref{arch}(a)), it includes the job description. Thus, all the generated job managers hold the DAG structure of the job.

When the job managers are in position, the pJM first decides the \textit{initial} task assignment among the job managers, and then the job managers \textit{cooperatively} schedule and generate tasks to execute (dot line in Fig.~\ref{arch}(b)) (\S\ref{Prades}). We call each sJM semi-active because it is not totally under control of the primary job manager, and it has freedom to determine the task assignment in its own cluster (dash line), to coordinate with other sJMs about task assignment, and to manage its compute resources according to resource utilization feedback (purple solid line) (\S\ref{af}).

After a task completes its computation on a partition of data, it reports to its job manager (pJM or sJM) about the output partition location.
The job manager collects the partition location information in its cluster, modifies the \textit{partitionList}, and then notifies other job managers to keep the consistency of partitionList. Besides the partitionList, \textsc{Houtu} includes jobId, stageId, executorList (the available executors from all data centers, including JMs and their associated roles), and taskMap (which task should be assigned by which JM) in a job's intermediate information (Fig.~\ref{arch}(b)). \textsc{Houtu} maintains a replication of the intermediate information in each data center.

Since the job managers operate synchronously, when the job completes, all of them will proactively release their resources as well as themselves to their data centers.
\subsubsection{Failure recovery}\label{Faultrecoverysec}
As stated in \S\ref{spot}, we focus on in this work the recovery of job-level failures, which is the failures of job managers.

When a semi-active job manager fails because of the unpredictable termination of its host, the primary job manager will notice it and then send a request through its local master to generate a new sJM in the remote data center (like steps 2a and 2b in Fig.~\ref{arch}(a)). This sJM starts with the original job description and the intermediate information in its cluster and recognises its role (as semi-active). It \textit{inherits} the containers belonging to the previous sJM, and \textit{continues} to operate as in normal.

If the primary fails, the semi-active job managers will elect a new primary using the consistent protocol (in Zookeeper). The new pJM updates and propagates the intermediate information about its role change. Next, the new primary \textit{continues} the process of the job, operates in normal and generates a new semi-active job manager to replace the failed pJM as above.

We assume that all the job managers would not fail simultaneously. Actually, it is of particular interest to study the problem which guarantees deterministic reliability of a job execution in the mixed environment (with reliable and unreliable instances) and minimizes the total monetary cost, however this is out of the scope of this work.
\section{Design}\label{design}
In this section, we first provide the problem statement of optimizing efficiency of jobs (\S\ref{problem}).
Next, we show how the JMs use resource utilization feedback to manage resources (step 3 in Fig.~\ref{arch}(a)) (\S\ref{af}). Then, we describe how the JMs schedule tasks within and between data centers (step 5 in Fig.~\ref{arch}(a)) (\S\ref{Prades}). Finally, we theoretically analyze the performance of the algorithms (\S\ref{analysis}).

\subsection{Problem statement}\label{problem}

\begin{figure}
  \centering
  \includegraphics[width=1.9in]{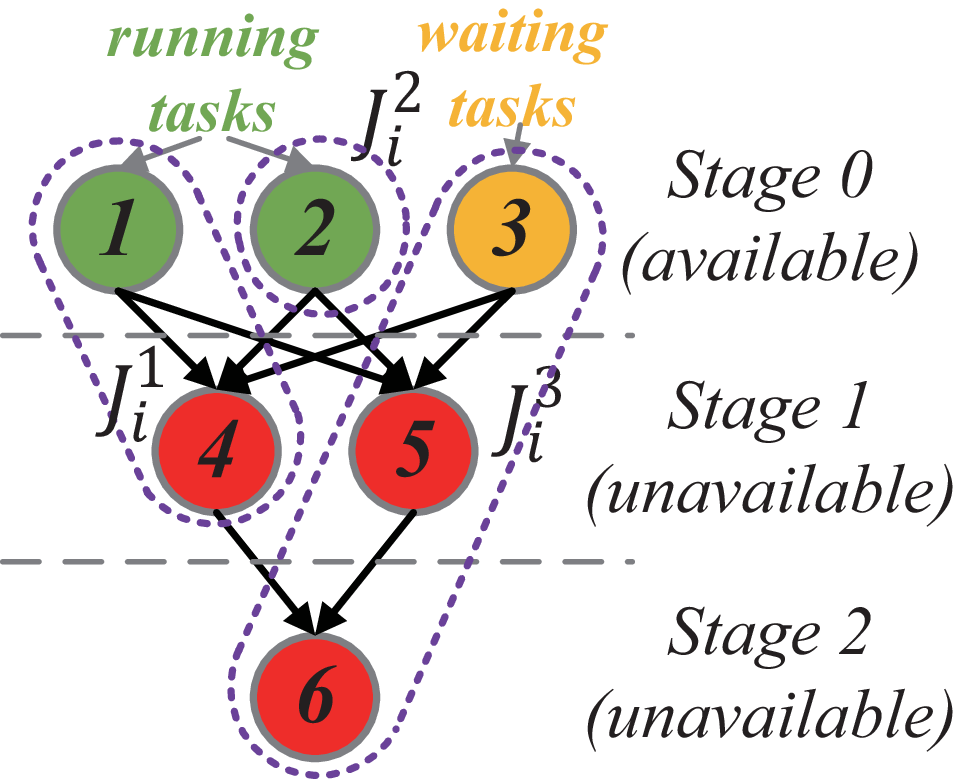}\\
  \caption{Example of a running DAG job $J_i$.}\label{job-info}
\end{figure}

Resources in \textsc{Houtu} are scheduled in terms of containers (corresponding to some \textit{fixed} amount of memory and cores). Instead of assuming the priori knowledge of \textit{complete} characteristics of jobs \cite{Tetris, Altruistic, GRAPHENE}, which restricts the types of workloads and incurs offline overheads, we rely on only \textit{partial} priori knowledge of a job (the knowledge from available stages). In the example of Fig.~\ref{job-info}, only the task information (including the input data locations, fine-grained
resource requirements, and process times) in \textit{Stage 0} is currently known, while the task information in \textit{Stage 1} and \textit{Stage 2} is currently unknown because they have not been
released yet. We consider that tasks in the same stage have identical characteristics, which conforms to the fact in practical systems as they perform the same computations on different
partitions of the input.

\begin{algorithm}[htb]
\caption{Af (applied by each job manager)}
\begin{algorithmic}[1]
  \Procedure{Af($d(q - 1)$, $a(q - 1)$, $u(q - 1)$)}{}
  \If{$q$ = 1}
  \State $d(q)$ $\gets$ 1
  \ElsIf{$u(q - 1)$ $<$ $\delta$ \textbf{and} \textit{no waiting tasks}}
  \State $d(q)$ $\gets$ $d(q - 1)$ / $\rho$  \hfill{//\textit{inefficient}}
  \ElsIf{$d(q - 1)$ $>$ $a(q - 1)$}
  \State $d(q)$ $\gets$ $d(q - 1)$         \hfill{//\textit{efficient and deprived}}
  \Else
  \State $d(q)$ $\gets$ $d(q - 1)$ $\cdot$ $\rho$ \hfill{//\textit{efficient and satisfied}}
  \EndIf
  \Return $d(q)$
  \EndProcedure
\end{algorithmic}
\end{algorithm}
In the scenario where multiple DAG jobs arrive and leave
online, we are interested in minimizing the makespan and average job response time.\footnote{The response time of a job is the duration time from its release to its completion.} Please refer to Appendix \ref{appenda} for the problem formulation. \textsc{Houtu} applies Af (Adaptive feedback algorithm) for each JM  to manage resources, and Parades in each JM to schedule tasks, which we will demonstrate in next two subsections, respectively.

\subsection{Resource management using Af}\label{af}

Resources in a data center are scheduled by the job scheduler to sub-jobs between \textbf{\textit{periods}}, each of equal time length $L$. We denote the \textbf{\textit{sub-job}} to the collection of tasks of a job that are executed in the same data center (and handled by the same job manager). Fig.~\ref{job-info} shows an example of sub-job partition with dot-line cycles.

For each sub-job $J_i^j$ of job $J_i$, its job manager (pJM or sJM) enforces Af (Algorithm 1) to determine the desire number of containers for next period $d(J_i^j, q)$ based on
its last period desire $d(J_i^j, q - 1)$, the last period allocation $a(J_i^j, q - 1)$, the last period resource utilization $u(J_i^j, q - 1)$ and waiting tasks.\footnote{We omit $J_i^j$ in Algorithm 1 for brevity.}
$u(J_i^j, q - 1)$  corresponds to the average resource utilization in period $q - 1$, and can be measured
by the monitoring mechanism.
\begin{algorithm}[htb]
\caption{Parades (applied by each job manager)}
\begin{algorithmic}[1]
  \Procedure{onUpdate($n$, $\delta$, $\tau$)}{}
  \State For each $t_{ij}$, increase $t_{ij}.wait$ by the time since last event \textit{UPDATE}; $cont$ $\gets$ true
  \If{ no waiting task}
  \State $t$ = \textsc{steal}($n$);
  \State $tlist$.add($t$); $n.\textit{free}$ $-$= $t.\textit{r}$; $cont$ $\gets$ false;
  \EndIf
  \While{$n.\textit{free}$ $>$ 0 \textbf{and} $cont$}
  \State $cont$ $\gets$ false;
  \If{there is a node-local task $t_{ij}$ on $n$ \textbf{and} $n.\textit{free}$ $\ge$ $t_{ij}.\textit{r}$}
  \State $t$ = $t_{ij}$ 
  \ElsIf{there is a rack-local task $t_{ik}$ on $n$ \textbf{and}  $n.\textit{free}$ $\ge$ $t_{ij}.\textit{r}$ \textbf{and} $t_{ik}.wait \ge \tau \cdot t_{ik}.p$ }
  \State $t$ = $t_{ik}$ 
  \ElsIf{there is a task $t_{il}$ with $t_{il}.wait \ge 2\tau \cdot t_{il}.p$ \textbf{and} $n.\textit{free}$ $\ge$ $1 - \delta$}
  \State $t$ = $t_{il}$ 
  \EndIf
  \State $tlist$.add($t$); $n.\textit{free}$ $-$= $t.\textit{r}$; $cont$ $\gets$ true;
  \EndWhile
  \Return $tlist$
  \EndProcedure
  \Procedure{onReceiveSteal($n$)}{}
  \State \Return \textsc{onUpdate($n$, $\delta$, $\tau$)}
  \EndProcedure
  \Procedure{steal($n$)}{}
  \For{each job manager of the same job}
  \State  $tlist$.add(\textsc{sendSteal($n$)})
  \EndFor
  \State \Return $tlist$
  \EndProcedure
\end{algorithmic}
\end{algorithm}

Consistent with \cite{ppopp}, we classify the period $q - 1$ as \textit{satisfied} versus \textit{deprived}.
Af compares the job's allocation $a(J_i^j, q - 1)$ with its desire $d(J_i^j, q - 1)$. The period is satisfied if
$a(J_i^j, q - 1) = d(J_i^j, q - 1)$, as the sub-job $J_i^j$ acquires as many containers as it requests from the job scheduler.
Otherwise, $a(J_i^j, q - 1) < d(J_i^j, q - 1)$, the period is deprived.
The classification of a period as \textit{efficient} versus \textit{inefficient} is more involved than that in \cite{ppopp}. Af uses a parameter $\delta$ as well as the presence of waiting task information.
The period is inefficient if the utilization $u(J_i^j, q - 1) < \delta$ \textit{and} there is no waiting task in period $q - 1$.
Otherwise the period is efficient.

If the period is inefficient, Af decreases the desire by a factor $\rho$.
If the period is efficient but deprived, it means
 that the sub-job efficiently used the resources it was allocated, but Af had requested more containers than
 the sub-job actually received from the job scheduler. It maintains the same desire in period $q$. If the period is efficient and satisfied,
 the sub-job efficiently used the resources that Af requests. Af assumes that the sub-job can use more containers and increases its desire by a factor $\rho$.
 In all three cases, Af allows Parades to assign multiple tasks to execute in a container.

\subsection{Task assignment using Parades}\label{Prades}

\textbf{Initial task assignment (applied by the primary job manager):} When a new stage of a DAG job becomes available, the primary job manager initially decides the fraction of tasks to place on each data center to be proportional to the amount of data  on the data center.

\textbf{Parades} (\textbf{Para}meterized \textbf{de}lay scheduling with work \textbf{s}tealing) is applied by each job manager after the initial assignment.
Parades is based on framework of the original delay scheduling algorithm \cite{DelaySched}, but extends it from two perspectives.
When a container updates its status, the algorithm adds the waiting time for each waiting task of the sub-job since the last event
\textit{UPDATE} happened (line 2), followed by the task
assignment procedure. Delay scheduling sets the waiting time thresholds for tasks as an invariant, while we modify the threshold for each task to be linearly dependent of its
processing time $p$ (which is \textit{known}), under the intuition of that ``long" tasks can tolerate a longer waiting time to acquire their preferred resources. On the other hand, if there is no waiting task, the job manager becomes a ``thief'' and tries to steal tasks from other ``victim'' job managers in the same job (line 4). Each victim job manager will handle this steal as a \textit{UPDATE} event (line 16).

Parades operates as follows in task
assignment procedure: It first checks whether there is a node-local task waiting, which means the container \textit{n} is on the same server as the task prefers. Assigning the
task to its preferred server which containing its input data helps in reducing data transmission over the network. We use $n.\textit{free}$ to denote
the free resources on container \textit{n}.  Secondly, the algorithm would check whether there
is a rack-local task for the \textit{n}, as the container shares the same rack as the task's preferred server. If the task has waited for more than the threshold
 time ($\tau \cdot t_{ij}.p$), and the container has enough free resources, we assign the task to the container. Finally, when a task has waited for long enough time
 ($2\tau \cdot t_{il}.p$), and $n.\textit{free}$ $\ge$ $1 - \delta$, we always allow the task could be assigned if possible.
 When $n.\textit{free}$ $\ge$ $1 - \delta$, the utilized resource of the container \textit{n} $< \delta$. We assume
 $t_{il}.\textit{r} + \delta \le 1$, for each $i, l$, as the upper bound for task resource requirement.

 Please refer to Table~\ref{notations} for the involved
 notations in our algorithms and their explanations.

\begin{table}
\caption{Explanations of notations.}\label{notations}
\centering
\begin{tabular}{c||c}
\cline{1-2}
\hline
\bfseries Notation & \bfseries Explanation \\
\hline \hline
$d(J_i^j, q)$ & $J_i^j$'s desire for period $q$ \\

$a(J_i^j, q)$ & $J_i^j$'s allocation for period $q$  \\

$u(J_i^j, q)$ & $J_i^j$'s resource utilization in period $q$  \\

$\delta$ & the utilization threshold parameter\\

$\rho$ & the resource adjustment parameter \\

$\tau$ & the task waiting time parameter  \\
\hline
\end{tabular}
\end{table}

\subsection{Analysis of Af + Parades}\label{analysis}
To prove the proposed algorithms guarantee efficient performance for online jobs, we settle the job scheduler employed in each data center as the fair scheduler \cite{FairSched, Max-min}, perhaps the most widely used job scheduler in both industry and academia. Once
there is a free resource, the fair scheduler always allocates it
to the job which currently occupies the fewest fraction of the
cluster resources, unless the job's requests have been satisfied.

We prove the following theorem about the competitive ratio of makespan.
Specifically, we \textit{extend} the very recent result \cite{xiaoda} about the efficiency of jobs scheduled by Af algorithm and parameterized delay (Pdelay) scheduling algorithm in a single data center.\footnote{We extend the Pdelay algorithm in \cite{xiaoda} with work stealing, which can \textit{only} accelerate task assignment and \textit{at most} delay tasks as much as in Pdelay algorithm.} Please see Appendix~\ref{makespan} for the proof sketch. We are still working on the provable efficiency about the average job response time.
\begin{thm}
When multiple geo-distributed DAG jobs arrive online and each data center applies fair job scheduler, the makespan of these jobs applying Af + Parades, is $O$(1)-competitive.
\end{thm}
\section{Implementation}\label{implementation}
We implement \textsc{Houtu} using Apache Spark \cite{Spark}, Hadoop YARN \cite{YARN} and Apache Zookeeper \cite{zk} as building blocks.
We make the following major changes:


\textbf{Monitor mechanism:} We estimate the dynamic resource availability on each container by adding a resource monitor process (in nodeManager component of YARN).
The monitor process reads resource usages (\textit{e.g.}, CPU, memory) from OS counters and reports them to its job manager.
Each job manager and its per-container monitors interact in an asynchronous manner to avoid overheads.

\textbf{Parameterized delay scheduling:} Based on the fact that tasks in a stage have similar resource requirements, we estimate the requirements
using the measured statistics from the first few
 executions of tasks in a stage. We continue to refine these estimations as more tasks have been
measured.
We estimate task processing time as the average processing time of all finished tasks in the same stage.
  We modify the original implementation of delay scheduling in Spark to take $\tau$ as a parameter read from the configuration file.

\textbf{How the job managers coordinate with each other?} As stated in \S\ref{normaloperation}, we use Zookeeper to synchronize JMs in the same job. Specifically, when the pJM determines the initial task assignment, it writes this information to taskMap (Fig.~\ref{arch}(b)). sJMs will notice this modification and begin their task assignment procedures using Parades (\S\ref{Prades}). If a job manager successfully steals a task from another, it also needs to modify the corresponding item in taskMap. After a task completes, it reports to its job manager about the output location, who will then propagate the location information in partitionList among other job managers.

\textbf{How a new job manager inherits the containers belonging to the failed one?} We modify YARN master to allow to grant tokens to the new generated job manager with the same jobId as the failed one. Then, the new job manager could use these tokens to access the corresponding containers.

\textbf{Af:} We continuously (per second) measure the container utilizations in a sub-job $J_i^j$ in a period $q$ of length $L$,
 and calculate the average at the end of the period. We acquire the desire number of containers for the next
 period $d(q+1)$ by Af (\S\ref{af}). If $d(q+1) \ge d(q)$, we directly update the desire and  push this
 new desire to the job scheduler. When $d(q+1) < d(q)$, the problem is involved, since we should decide \textit{Which} containers should be killed, and \textit{when} the kill should be performed? We aggressively kill the several containers which firstly become free. We add the control information through the job manager in Spark to negotiate resources with YARN master.

\section{Experimental Evaluation}\label{evaluation}
In this section, we first present the methodology in conducting our experiments (\S\ref{Methodology}). Then, we show the efficient job performance \textsc{Houtu} guarantees in both normal operation and changeable environment (\S\ref{Jobperformance}), and analyze the monetary costs of \textsc{Houtu} and other deployments when running the same workloads (\S\ref{Costanalysis}).
Finally, we verify the ability of recovering of job manager failures in \textsc{Houtu} (\S\ref{Faultrecovery}) and measure the overheads that it introduces in detail (\S\ref{overheadsec}).
\subsection{Methodology}\label{Methodology}
\begin{figure}
  \centering
  \includegraphics[width=2.0in]{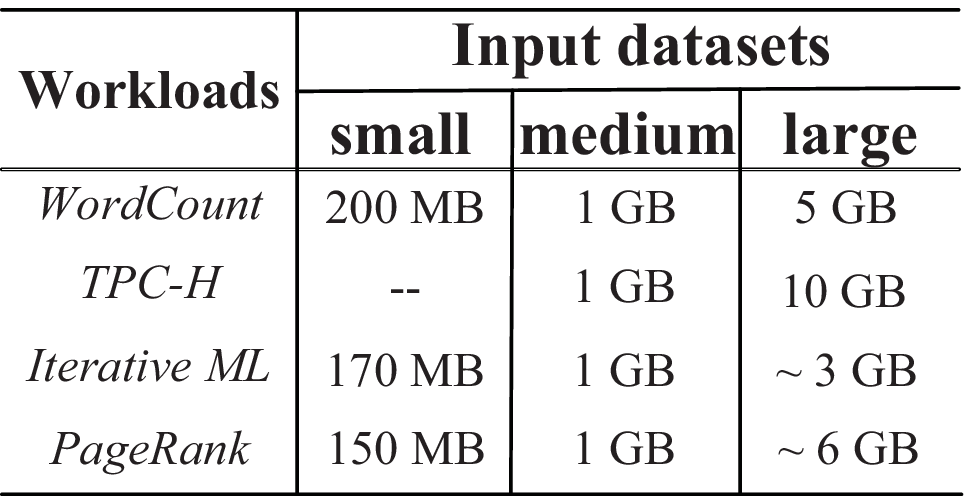}\\
  \caption{Input sizes for four workloads.}\label{workloads}
\end{figure}
\textbf{Testbed:} We deploy \textsc{Houtu} to 20 machines spread across four AliCloud regions as we show in \S\ref{changeable}. In each region, we start five machines of type \texttt{n4.xlarge} or \texttt{n1.large}, depending on their availability. Both types of instances have 4 CPU cores, 8GB RAM and run 64-bit Ubuntu 16.04. In each region, we choose one On-demand instance as the master and four Spot instances as workers.

\textbf{Workload:} We use workloads for our evaluation including \textit{WordCount}, \textit{TPC-H benchmark}, \textit{Iterative machine learning} and \textit{PageRank}.
For each workload, the variation in input sizes is based on real workloads from Yahoo! and Facebook,
in scale with our deployment (Fig.~\ref{workloads}).
For the job distribution, we set 46\%, 40\% and 14\% of jobs are with small, medium and large input sizes respectively,
which also conforms to realistic job distribution \cite{YARN}. For \textit{TPC-H benchmark}, we place in each data center two tables, while for other three workloads, we evenly partition
the input across four data centers.

\textbf{Baselines:} We evaluate the effectiveness of \textsc{Houtu} by evaluating four main types of systems/deployments: (1) the centralized Spark on YARN system with built-in static resource scheduling (cent-stat); (2) the centralized Spark on YARN system with state-of-the-art dynamic resource scheduling (cent-dyna) \cite{xiaoda}; (3) \textsc{Houtu}, decentralized architecture with Af + Parades; (4) decentralized architecture with static resource scheduling (decent-stat).
\begin{figure}[!t]
\centering
\subfigure[CDF of job response time.]
 {\includegraphics[width=1.6in]{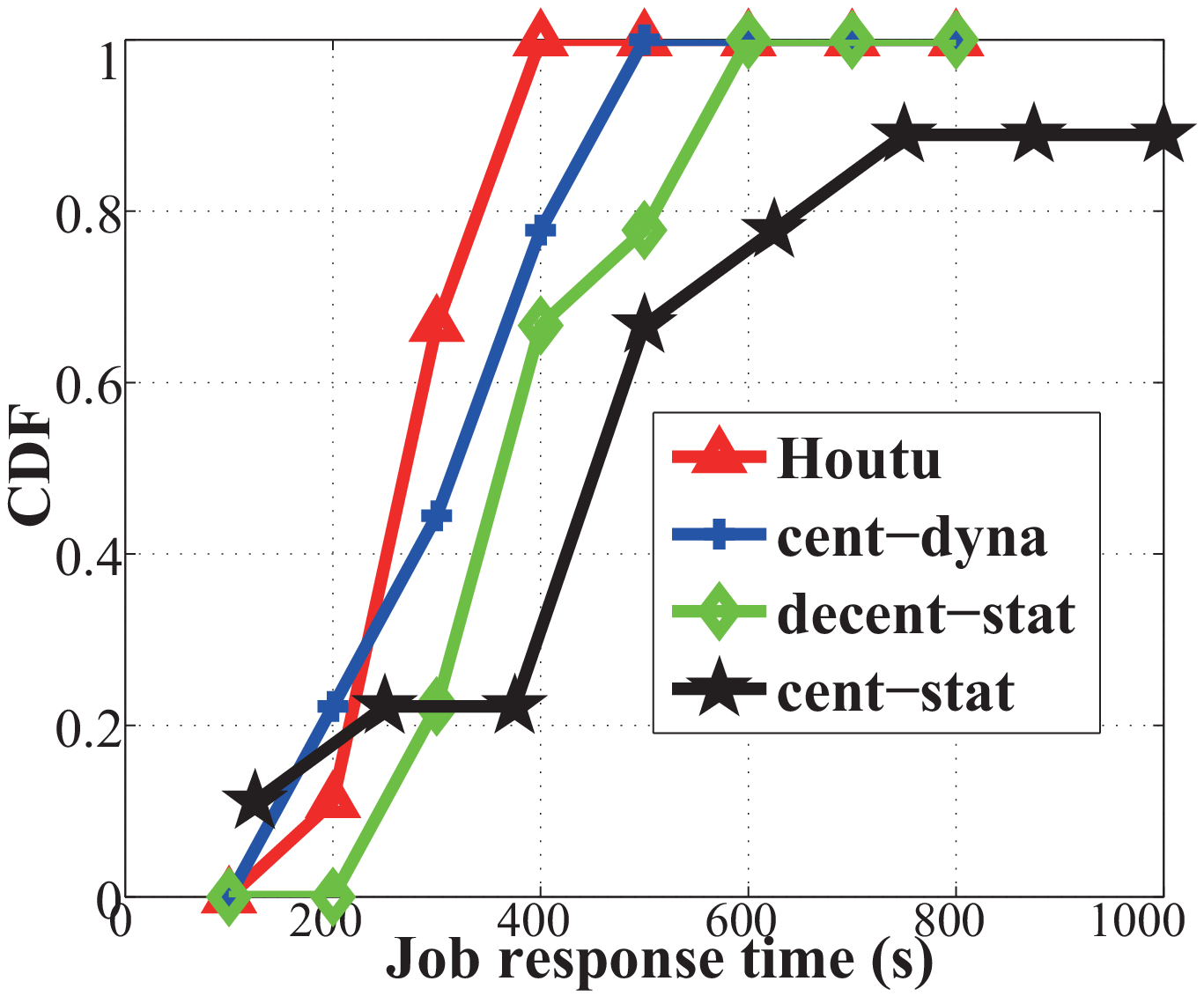}}
\subfigure[average job response time and makespan.]
{\includegraphics[width=1.4in]{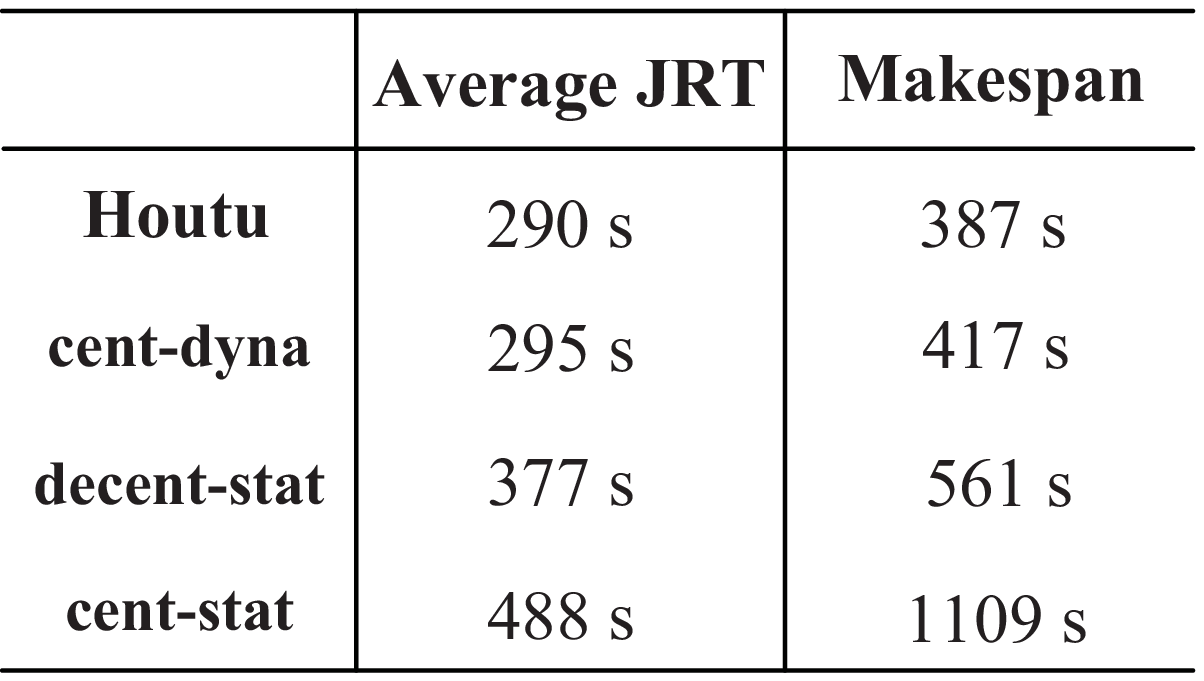}}
\caption{Job performance in four deployments.}\label{jrt}
\end{figure}

\textbf{Metrics:} We use average job response time and makespan to evaluate the effectiveness of jobs which arrive in an online manner. We also care about the monetary cost of running these jobs, compared with the deployment using total reliable (On-demand) instances. Finally, we are interested in job response times when facing failures.

\subsection{Job performance}\label{Jobperformance}

\begin{figure*}
\centering
\subfigure[The normal job execution in \textsc{Houtu}]{
\begin{minipage}[b]{0.3\textwidth}
\includegraphics[width=1\textwidth]{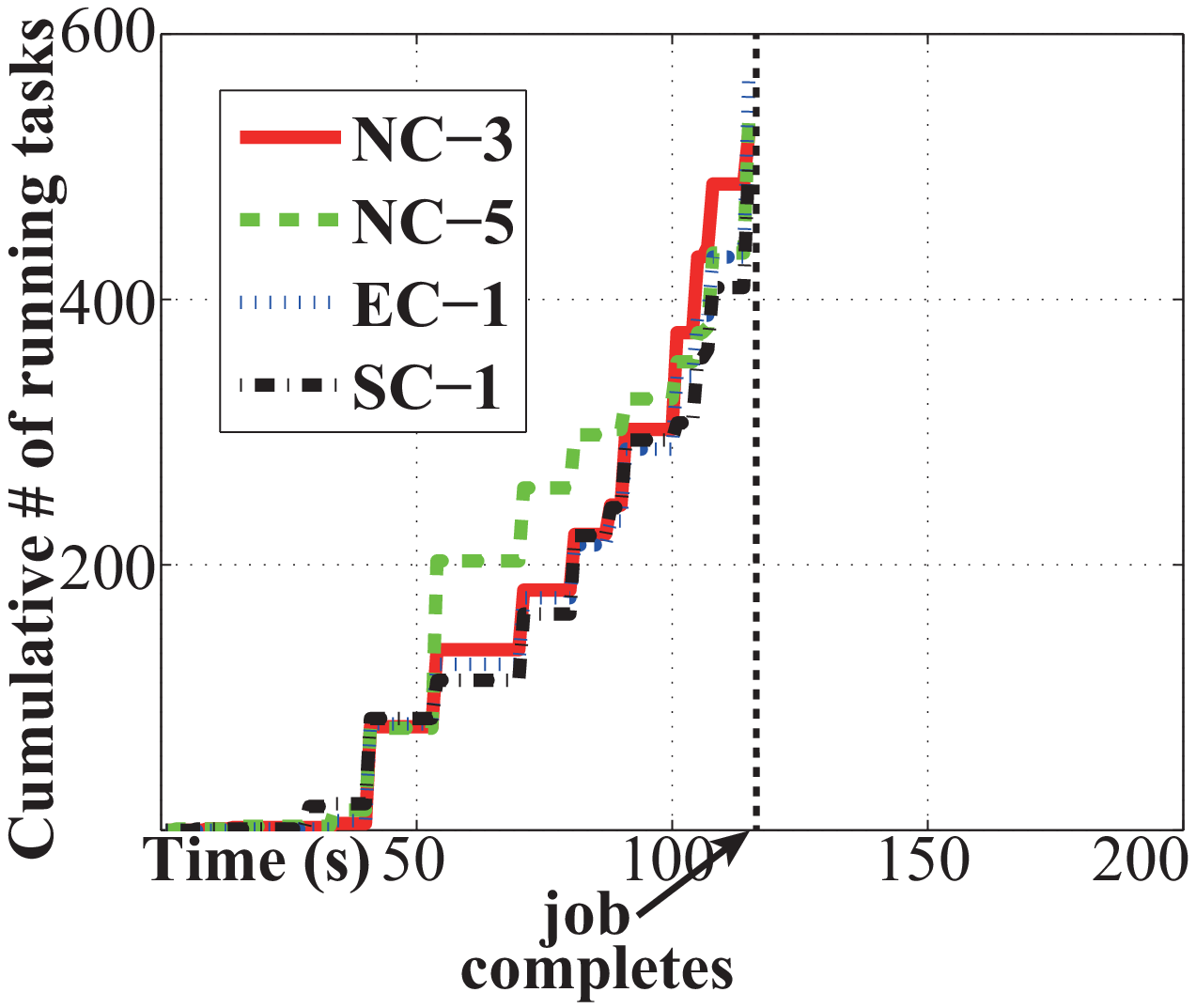}
\end{minipage}
}
\subfigure[The normal job execution in \textsc{Houtu} when we inject workloads]{
\begin{minipage}[b]{0.3\textwidth}
\includegraphics[width=1\textwidth]{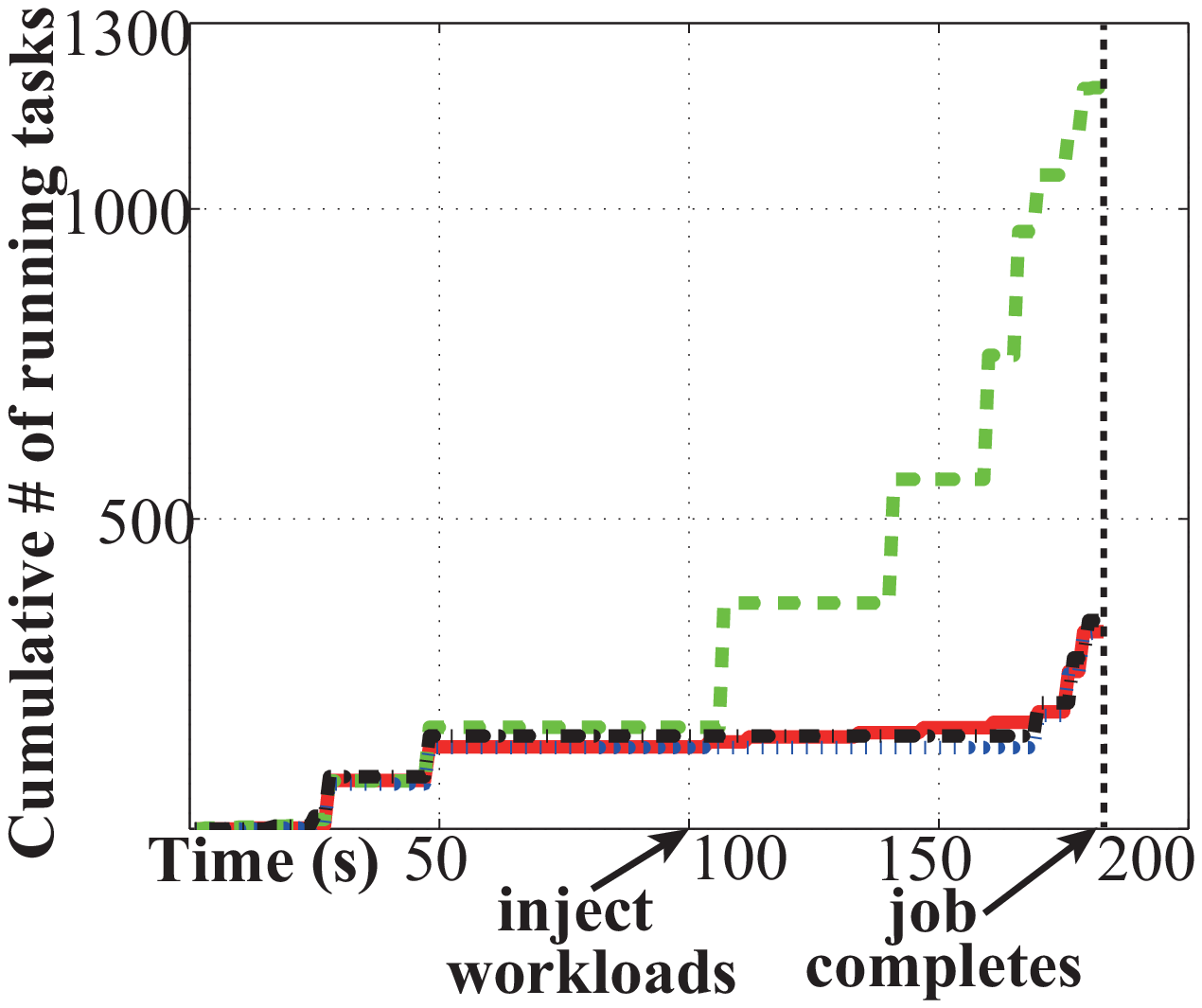}
\end{minipage}
}
\subfigure[The job execution without work stealing mechanism when we inject workloads]{
\begin{minipage}[b]{0.3\textwidth}
\includegraphics[width=1\textwidth]{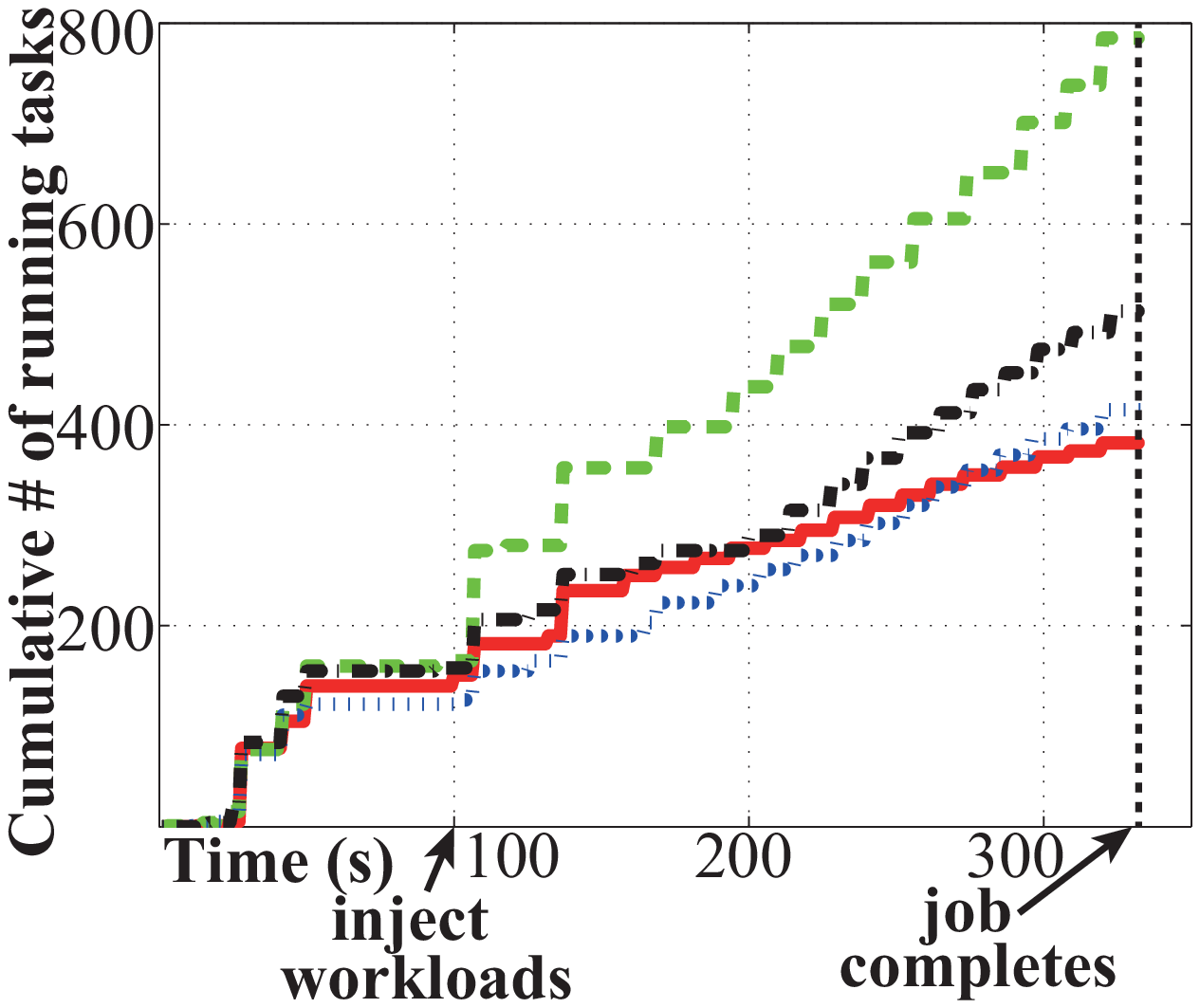}
\end{minipage}
}
\caption{The cumulative running tasks of a job execution in different scenarios and mechanisms.}\label{runningtasks}
\end{figure*}
We use the workloads stated before, and set the job submission time following an exponential distribution
with mean interval as 60 seconds.

Fig.~\ref{jrt} shows the job performance in our four different deployments. First, we find that \textsc{Houtu} has approximate performance compared with the centralized architecture with start-of-the-art dynamic scheduling mechanism. This approximation is due to that we allow job managers in a job to \textit{share} resources across data centers by work stealing (Parades). Second, When compared with the decentralized architecture with static scheduling algorithm, \textsc{Houtu} has $29\%$ improvement in terms of average job response time, and $31\%$ improvement in terms of makespan. This gain comes from the use of adaptively scheduling mechanism based on utilization feedback (Af).

To further demonstrate that \textsc{Houtu} guarantees efficient job performance in a changeable environment, we intentionally inject workloads to consume spare resources in data centers and see how a job reacts to this variation. Fig.~\ref{runningtasks} shows the cumulative running tasks of a job execution in different scenarios and mechanisms. In Fig.~\ref{runningtasks}(a), a job executes normally and completes at time $115$. While in Fig.~\ref{runningtasks}(b) and Fig.~\ref{runningtasks}(c), we inject workloads into three data centers NC-3, EC-1 and SC-1 to \textit{use up almost all} spare resources in these data centers at time $100$ after a job submission. Fig.~\ref{runningtasks}(b) demonstrates that work stealing mechanism ensures that the job manager in NC-5 \textit{gradually} steals tasks from the other resource-tense data centers as the new stages of the DAG job become available. However, without work stealing, the pJM assigns tasks only according to the data distribution (initial assignment), which then leads to that the sJMs in resource-tense data centers would \textit{queue} the tasks to be executed. As shown in Fig.~\ref{runningtasks}(c), the queueing delays the job.  Job response times in the last scenarios are $183$ and $333$ seconds, respectively.

\subsection{Cost analysis}\label{Costanalysis}

In this subsection, we configure the centralized architecture with On-demand instances, while we keep the decentralized architecture configuration with Spot instances (except the masters). We use the same workloads as in Fig.~\ref{jrt}, and calculate the monetary costs in different deployments. Costs are divided into machine cost and data transfer cost across different data centers\footnote{In AliCloud \cite{aliyunprice}, the price of data transfer across data centers is 0.13\$/GB, while it is free to transfer data within a data center.}.

\begin{figure}
  \centering
  \includegraphics[width=2.2in]{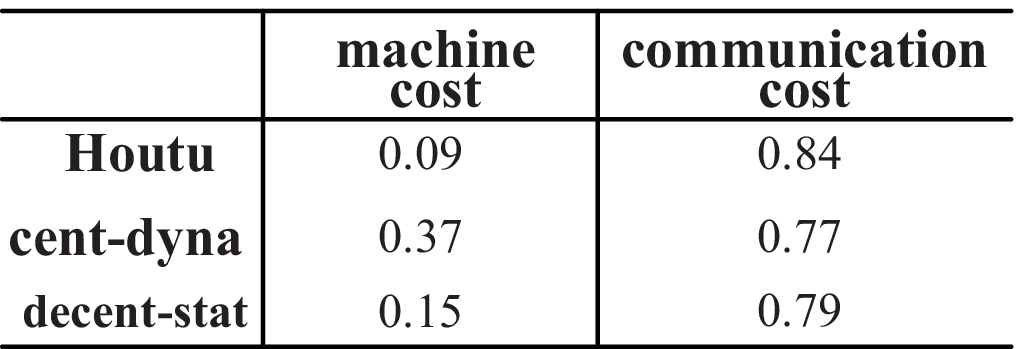}\\
  \caption{Normalized cost of different deployments.}\label{money}
\end{figure}

Fig.~\ref{money} shows two types of costs in different deployments normalized with the cost in cent-stat. First, we observe \textsc{Houtu} is very effective in reducing the machine cost of running geo-distributed jobs, which is $90\%$ cheaper than the cost in cent-stat. Not surprisingly, the major cost saving comes from the use of Spot instances. Second, \textsc{Houtu} has fewer data transfer compared with centralized architectures. This is because centralized architectures do not distinguish machines in different data centers; while \textsc{Houtu} differentiates task assignment within a data center and between data centers, and a task steal happens \textit{only after} the thief job manager finishes its own tasks. \textsc{Houtu} saves about $20\%$ communication cost compared with cent-stat.

\subsection{Failure recovery}\label{Faultrecovery}
\begin{figure*}
\centering
\subfigure[Recovery of a pJM failure in \textsc{Houtu}]{
\begin{minipage}[b]{0.3\textwidth}
\includegraphics[width=1\textwidth]{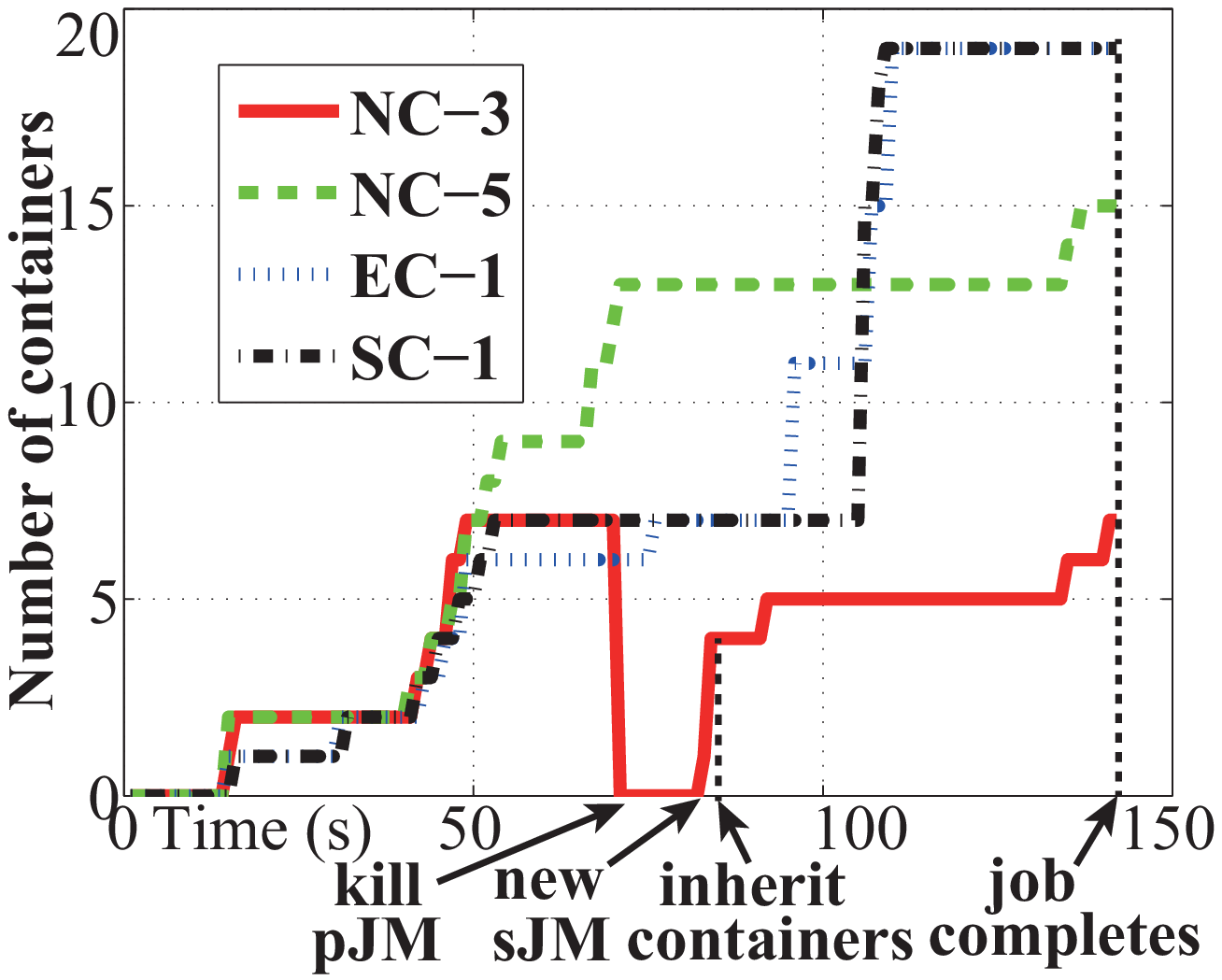}
\end{minipage}
}
\subfigure[Recovery of a sJM failure in \textsc{Houtu}]{
\begin{minipage}[b]{0.3\textwidth}
\includegraphics[width=1\textwidth]{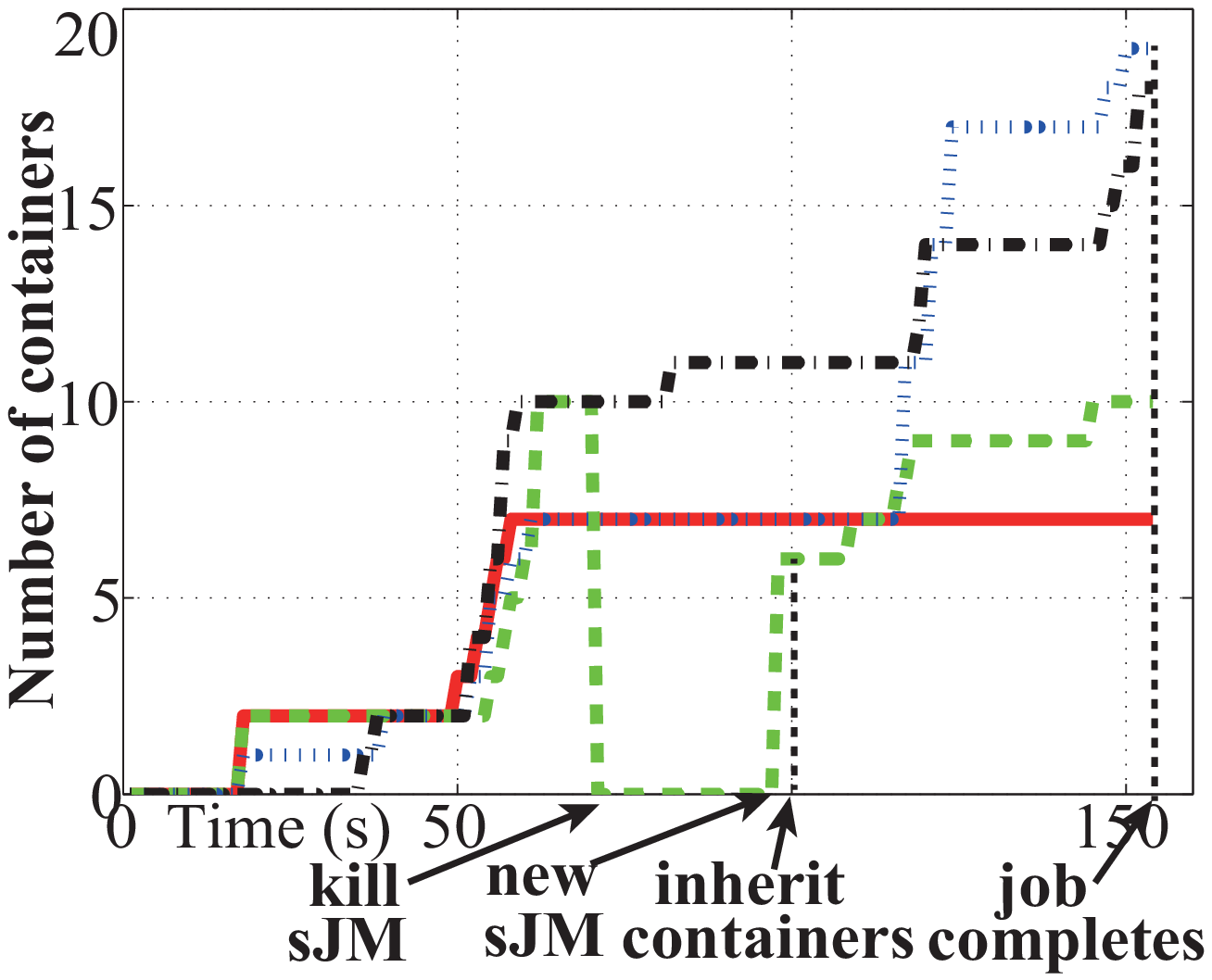}
\end{minipage}
}
\subfigure[A job manager failure in cent-dyna]{
\begin{minipage}[b]{0.3\textwidth}
\includegraphics[width=1\textwidth]{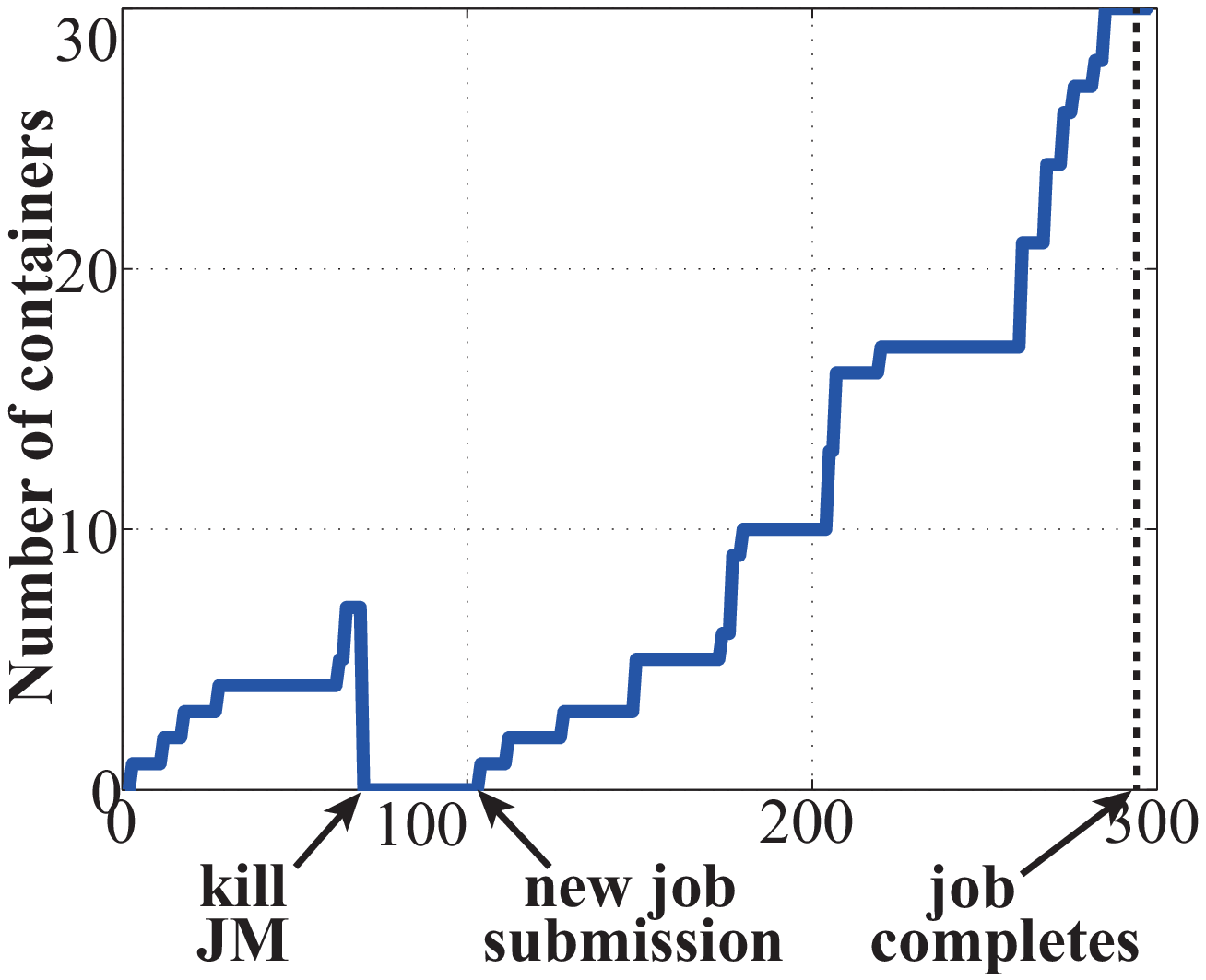}
\end{minipage}
}
\caption{The job failure recovery in \textsc{Houtu} and the centralized architecture}\label{failover}
\end{figure*}
One of our major design considerations of \textsc{Houtu} is to ensure that a job could recover from a failure due to the unreliable environment and continue to execute. To understand the
effectiveness of our proposed mechanism, we respectively run a job in \textsc{Houtu} and cent-dyna, and we \textit{manually} terminate the host (VM) where the job manager resides at $70$ seconds after its submission.

We count the number of containers belonging to the job. Fig.~\ref{failover} shows the process of the job execution experiencing a job manager failure. In Fig.~\ref{failover}(a), we kill the VM which hosts the pJM, and after $10$ seconds we see a new \textit{sJM} replaces the failed pJM.\footnote{A new pJM is first elected and then the new pJM tells the master where the former pJM resided to generate a new sJM (\S\ref{Faultrecoverysec}).} The sJM then inherits the old containers and continues its work. While in Fig.~\ref{failover}(b), we kill a sJM and see the similar process. The interval time is always lower than $20$ seconds in our extensive experiments. The job response times in two scenarios are $147$ seconds and $154$ seconds, respectively. However, in the centralized architecture, the failure of a job manager leads to the resubmission of the job, which wastes the previous computations. The job response time is $299$ seconds in the last case, which is significantly longer than the times in two executions in \textsc{Houtu}.
\subsection{Overhead}\label{overheadsec}
\begin{figure}[!t]
\centering
\subfigure[Intermediate info. size of four workloads on \textit{large} input.]
 {\includegraphics[width=1.5in]{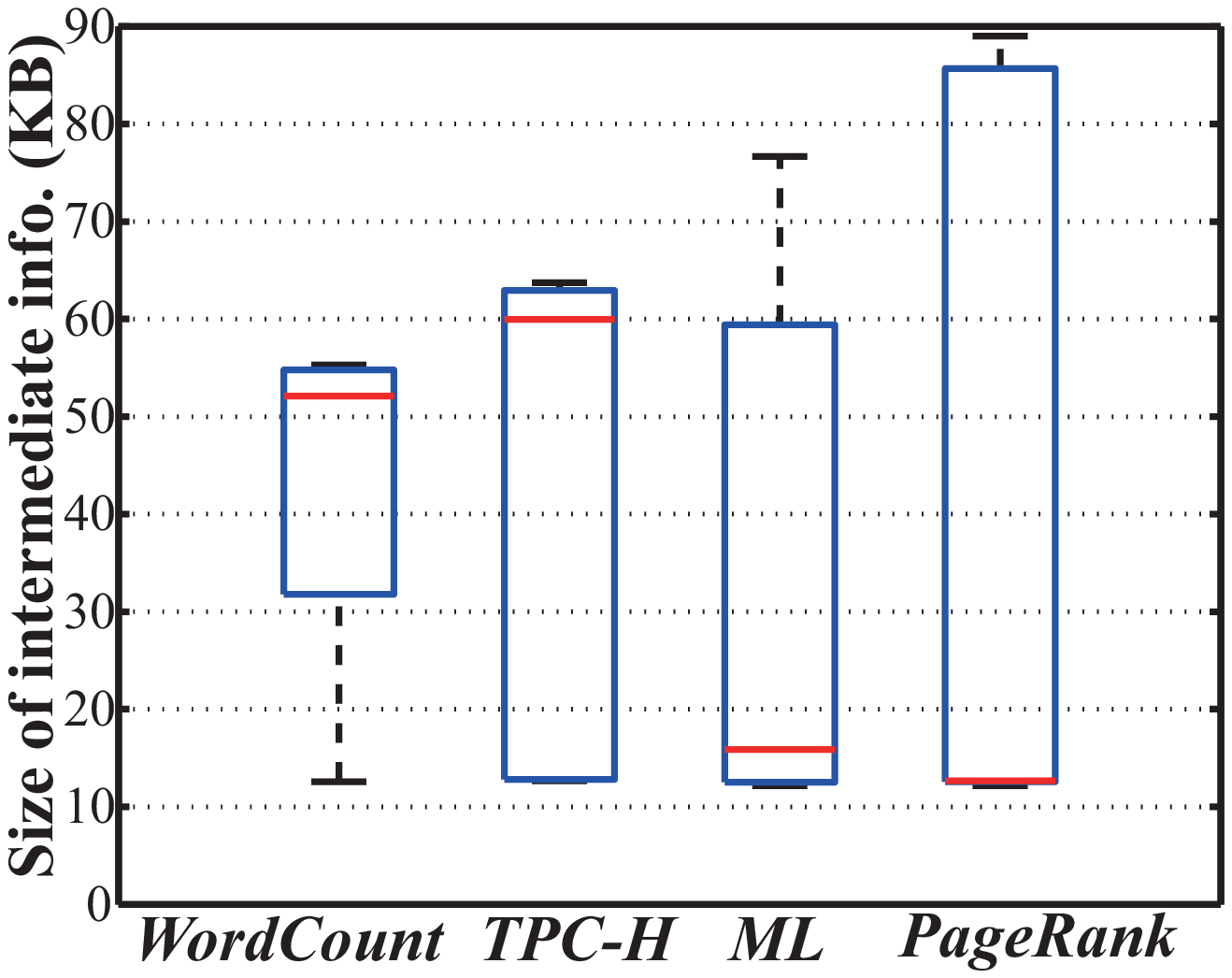}}
\subfigure[Time cost of different mechanisms in \textsc{Houtu}.]
{\includegraphics[width=1.5in]{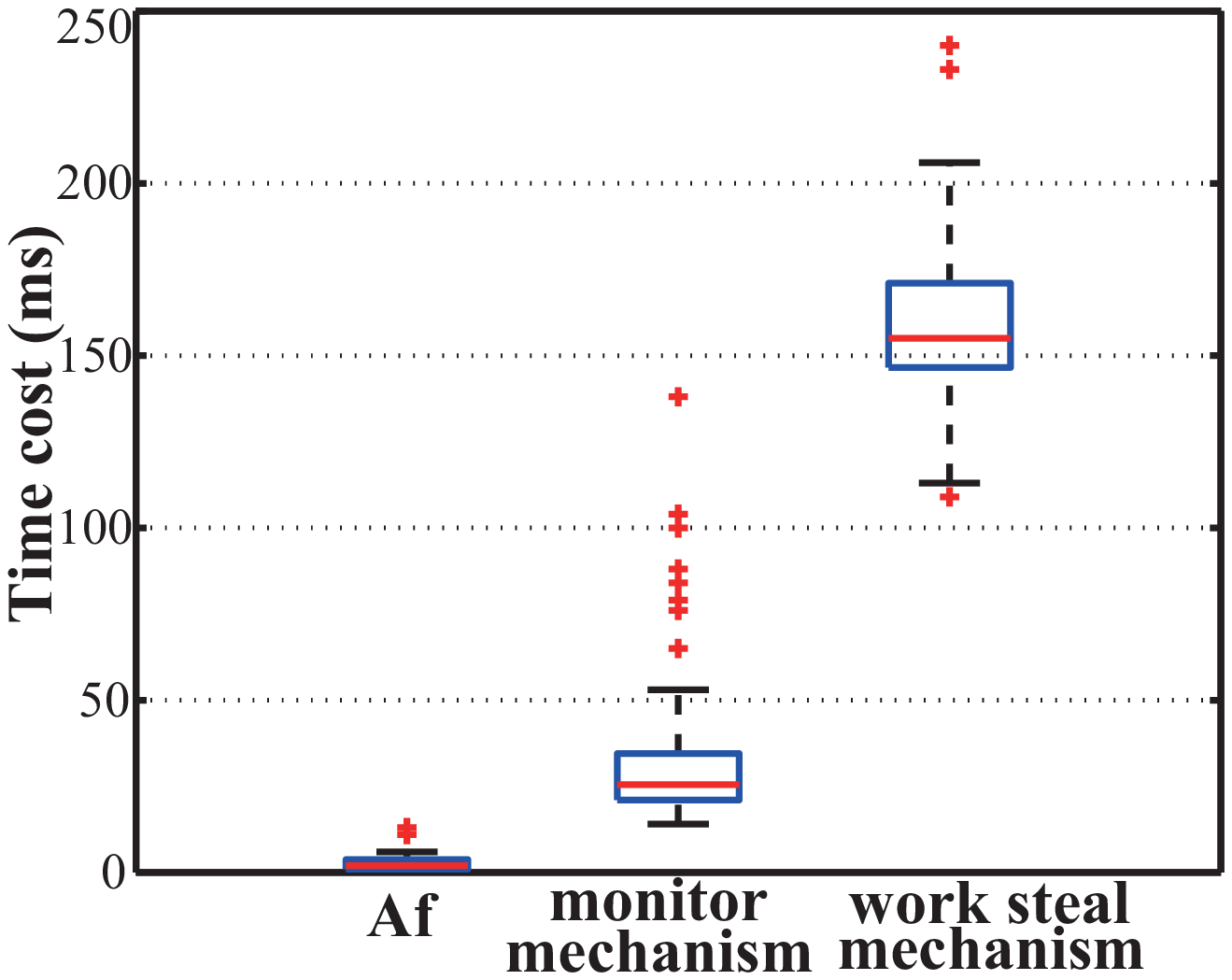}}
\caption{Intermediate information size and time cost.}\label{overheads}
\end{figure}
We measure overheads of \textsc{Houtu} from two perspectives.

First, we collect the intermediate information of jobs from four workloads on large input datasets, and measure their sizes during their executions. Fig.~\ref{overheads}(a) plots the 25th percentile, median and 75th percentile sizes for each workload in the corresponding box. We find the average sizes for the four workloads are $43.1$ KB, $43.4$ KB, $37.8$ KB and $30.8$ KB, respectively, which are small enough to use Zookeeper to keep them consistent.

Second, we measure the time costs of mechanisms that \textsc{Houtu} introduces. For the Af overhead, it just maintains the update operation and incurs negligible costs. Compared to the default implementation in YARN, we add the monitoring mechanism in each container process, which has moderate overhead. As a job manager incurs transmission delay in work stealing, we find the average delay of the steal message transmissions is $163.5$ ms across different system loads, which is also acceptable.
\section{Related Work}

\textbf{Wide-area data analytics:} Prior work establishes the emerging problem of analyzing the globally-generated data in data analytics systems \cite{Regulatory, Hung, Pu, CLARINET, Lube, Gaia, baochun}. These works show promising WAN bandwidth reduction and job performance improvement. SWAG \cite{Hung} adjusts the order of jobs across data centers to reduce job completion times. Iridium \cite{Pu} optimizes data and task placement to reduce query response
times and WAN usage. Clarinet \cite{CLARINET} pushes wide-area network awareness to the
query planner, and selects a query execution plan before the query begins. The proposed solutions work in the centralized architecture and assume the WAN bandwidth as constant, however, these may not conform to the practical scenario due to our argument in \S\ref{background}. In contrast, we focus on the design of the decentralized geo-distributed data analytics architecture and requires no modification to the current job descriptions.

\textbf{Scheduling in a single data analytics system:} Data-locality is a primary goal when scheduling tasks within a job. Delay scheduling \cite{DelaySched}, Quincy \cite{Quincy} and Corral \cite{corral} try to improve the locality of individual
tasks by scheduling them close to their input data. Fairness-quality tradeoff between multiple jobs is another goal. Carbyne \cite{Altruistic} and Graphene \cite{GRAPHENE} improve cluster utilizations and performances while allowing a little unfairness among jobs. Most of these systems rely on the priori knowledge of
DAG job characteristics. Instead, we use utilization feedback to dynamically adjust scheduling decisions with only partial priori knowledge. Further, we extend this mechanism in the context of geo-distributed data centers and allow job managers to cooperate in scheduling tasks.

\textbf{Fault-tolerance for jobs in data analytics:} In current systems like MapReduce \cite{mapreduce}, Dryad \cite{Dryad} and Spark \cite{Spark}, each job manager tracks the execution time of every task, and reschedules a copy task when the execution time exceeding a threshold (straggler). At the level of jobs, the cluster (job scheduler) will resubmit a job when its reports are absent for a while. The resubmitted job starts its execution from scratch, wasting the previous computations. In the relevant grid computing, fault-tolerance of jobs is achieved by checkpointing, which is the collection of process context states \cite{nguyen, jpdcLee}. The process context states are stored periodically on a stable storage, which is not applicable in the data analytics systems due to the overhead for each job manager collecting the real-time task states and then persisting them. We include the output location for each task (partitionList) instead of its context state in its intermediate information, which is effective and incurs acceptable overheads as evidenced in our experiments.
\section{Conclusion}
We introduce \textsc{Houtu}, a new data analytics system that is designed to support analytics jobs on globally-generated data with respect to the practical constraints, without any need to change the jobs. \textsc{Houtu} provides a job manager for a job in each data center, ensuring the reliability of its execution. We present the strategy for each JM to independently manage resources without complete priori knowledge of jobs, and the mechanism for each JM to assign tasks which can adjust its decisions according to the changeable environment. We experimentally verify \textsc{Houtu}'s functionalities to guarantee reliable and efficient job executions. We conclude that  \textsc{Houtu} is a practical and effective system to enable constrained globally-distributed analytics jobs.

\appendix

\section{Problem Formulation}\label{appenda}

Suppose there is a set of jobs $\mathcal{J} = \{J_1, J_2, ..., J_{|\mathcal{J}|}\}$ to be
scheduled on a set of containers $\mathcal{P} = \{P_1, P_2, ..., P_{|\mathcal{P}|}\}$ from all data centers.
These containers are different since they reside in different servers (and different data centers) containing different input data for jobs.
Time is discretized into $\textit{\textbf{scheduling periods}}$ of equal length $L$,
where each period $q$ includes the interval [$L\cdot q$, $L\cdot (q+1) - 1$]. $L$ is a configurable system parameter.

We model a job $J_i$ as a DAG. Each vertex of the DAG represents a task and each edge
represents a dependency between the two tasks. Each task in a job prefers a unique subset of $\mathcal{P}$, as the containers in the subset
store the input data for the task. For each task $t_{ij}\in J_i$, we denote by $t_{ij}.\textit{r}$
 to be the peak requirements. We assume
$0 \le t_{ij}.\textit{r} \le 1$, normalized by the container capacity. We also assume $t_{ij}.\textit{r} \ge \theta$,
where $\theta > 0$, \textit{i.e.}, a task must consume some amount of resources. We associate $t_{ij}.p$ to be the processing time of task $t_{ij}$.
Furthermore, the $\textit{\textbf{work}}$ of a job $J_i$ is defined as $T_1(J_i) =  \sum_{t_{ij}\in J_i} t_{ij}.\textit{r}\cdot t_{ij}.p$.
 The release time $r(J_i)$ is the time at which the job $J_i$ is submitted.
 A task is called in the \textit{waiting} state when its predecessor tasks have all completed and itself has not been scheduled yet.

 The \textit{\textbf{sub-job}} $J_i^j$ of $J_i$ corresponds to a collection of tasks executing in the data center $j$.
 Each job manager handles the task executions of a sub-job in the job manager's data center. The job managers of a job are oblivious to the further characteristics of the unfolding DAG.

\begin{defn}
The \textbf{makespan} of a job set $\mathcal{J}$ is the time taken to complete all the jobs in $\mathcal{J}$, that is, \emph{T($\mathcal{J}$)} = $\max_{J_i\in \mathcal{J}}$ \emph{T}($J_i$), where \emph{T}($J_i$) is the completion time of job $J_i$.
\end{defn}

\begin{defn}
The \textbf{average response time} of a job set $\mathcal{J}$ is given by $\frac{1}{|\mathcal{J}|}$ $\sum_{J_i\in \mathcal{J}} (\emph{T}(J_i) - r(J_i))$.
\end{defn}

The job scheduler of a data center and a job manager interact as follows. The job scheduler reallocates resources between scheduling periods.
At the end of period $q - 1$, the job manager of sub-job $J_i^j$ determines its desire $d(J_i^j, q)$, which is the number of containers
$J_i^j$ wants for period $q$. Collecting the desires from all running sub-jobs, the job scheduler decides allocation
$a(J_i^j, q)$ for each sub-job $J_i^j$ (with $a(J_i^j, q) \le d(J_i^j, q)$). Once a job is allocated containers, the job manager further
schedules its tasks. And the allocation does not change during the period.

Given a job set $\mathcal{J}$ and container set from all data centers $\mathcal{P}$, we seek for a combination of a job scheduler (how to allocate resources to sub-jobs), and job managers within each job (how to request resources and how to assign tasks to the given resources),
which minimizes makespan and average response time of $\mathcal{J}$, while satisfying the task locality preferences.

\section{Efficiency of the Makespan}\label{makespan}
We first state a theorem from \cite{xiaoda} and then use it to prove the efficiency of makespan in the context of geo-distributed DAG jobs running in multiple data centers.
\begin{thm}
\cite{xiaoda} In a single data center with  container set $\mathcal{P}$, which applies fair job scheduler, when DAG jobs $\mathcal{J}$ running in it with each applying Adaptive feedback algorithm to request resources and parameterized delay scheduling to assign tasks, the makespan of these jobs is
\begin{eqnarray*}
\textit{\emph{T}}(\mathcal{J}) &\le& (\frac{2}{1 - \delta} + \frac{1 + \rho}{\delta} + \frac{2\tau}{\theta}) \frac{T_1(\mathcal{J})}{|\mathcal{P}|} \\
&+& L\log_{\rho} |\mathcal{P}| + 2L\ .
\end{eqnarray*}
\end{thm}

Assume there are $k$ data centers, the sub-job set executing in data center $j$ is $\mathcal{J}^j$ and there are $|\mathcal{P}_j|$ containers in date center $j$. In the $J_i$ example of Fig.~\ref{job-info}, $J_i^1 \in \mathcal{J}^1$, $J_i^2 \in \mathcal{J}^2$, and $J_i^3 \in \mathcal{J}^3$. Denote $c_i = \frac{1}{|\mathcal{P}_i|}(\frac{2}{1 - \delta} + \frac{1 + \rho}{\delta} + \frac{2\tau}{\theta})$ and $d_i = L\log_{\rho} |\mathcal{P}_i| + 2L$. By directly applying theorem 2, we have for each $i$,
\begin{eqnarray*}
\textit{\emph{T}}(\mathcal{J}^i) \le c_i\cdot T_1(\mathcal{J}^i) + d_i\ .
\end{eqnarray*}
Sum them up, we have
\begin{eqnarray*}
\sum_{i=1}^{k}\textit{\emph{T}}(\mathcal{J}^i) &\le& c_{max}\cdot \sum_{i=1}^{k}T_1(\mathcal{J}^i) + \sum_{i=1}^{k}d_i\ \\
&=&c_{max} \cdot T_1(\mathcal{J}) + \sum_{i=1}^{k}d_i \\
&=&c_{max}|\mathcal{P}|\cdot \frac{T_1(\mathcal{J})}{|\mathcal{P}|} + \sum_{i=1}^{k}d_i\ ,
\end{eqnarray*}
in which $c_{max}$ is the max of $c_i$ and the first equality is due to the definition of work. According to the fact $\sum_{i=1}^{k}\textit{\emph{T}}(\mathcal{J}^i) \ge \textit{\emph{T}}(\mathcal{J})$, we have
\begin{eqnarray*}
\textit{\emph{T}}(\mathcal{J}) \le c_{max}|\mathcal{P}|\cdot \frac{T_1(\mathcal{J})}{|\mathcal{P}|} + \sum_{i=1}^{k}d_i\ .
\end{eqnarray*}
Since $\frac{T_1(\mathcal{J})}{|\mathcal{P}|}$ is a lower bound of $\textit{\emph{T}}^{*}(\mathcal{J})$ due to \cite{lowerbound}, and the number of available containers in all data centers $|\mathcal{P}|$  is constant once the
system is well configured, we complete the proof of theorem 1.

\bibliography{atc}
\bibliographystyle{abbrv}

\end{document}